\documentclass[floats,floatfix,showpacs,amssymb,prd,twocolumn,superscriptaddress,nofootinbib]{revtex4-1}

\newlength{\figw} 
\usepackage{graphicx,epsf, epsfig, amssymb}
\usepackage{bm}
\usepackage{longtable,tabularx}
\usepackage{color}
\usepackage[breaklinks]{hyperref}
\usepackage{amsfonts,amsmath,amssymb,mathrsfs,gensymb}
\usepackage{natbib}
\usepackage{prd_macros}
\usepackage{array}
\usepackage{multirow}
\usepackage{rotating,array}
\usepackage{graphicx}
\usepackage[normalem]{ulem}
\usepackage[caption=false]{subfig}

\graphicspath{ {images/} }

\newcommand\optional[1]{}
\def\be{\begin{equation}}
\def\ee{\end{equation}}
\def\beq{\begin{eqnarray}}
\def\eeq{\end{eqnarray}}
\usepackage{comment}

\begin{document}
\title{A unified treatment of tidal disruption by Schwarzschild black holes}

\author{Juan Servin}
\email{jes105020@utdallas.edu}
\affiliation{Department of Physics, The University of Texas at Dallas, Richardson, TX 75080, USA }

\author{Michael Kesden}
\email{kesden@utdallas.edu}
\affiliation{Department of Physics, The University of Texas at Dallas, Richardson, TX 75080, USA }

\date{\today}

\begin{abstract}
Stars on orbits with pericenters sufficiently close to the supermassive black hole at the center of their host galaxy can be
ripped apart by tidal stresses.  Some of the resulting stellar debris becomes more tightly bound to the hole and can
potentially produce an observable flare called a tidal-disruption event (TDE).  We provide a self-consistent, unified treatment
of TDEs by non-spinning (Schwarzschild) black holes, investigating several effects of general relativity including changes to
the boundary in phase space that defines the loss-cone orbits on which stars are tidally disrupted or captured.   TDE rates
decrease rapidly at large black-hole masses due to direct stellar capture, but this effect is slightly countered by the widening
of the loss cone due to the stronger tidal fields in general relativity.  We provide a new mapping procedure that translates
between Newtonian gravity and general relativity, allowing us to better compare predictions in both gravitational theories.
Partial tidal disruptions in relativity will strip more material from the star and produce more tightly bound debris than in
Newtonian gravity for a stellar orbit with the same angular momentum.  However, for deep encounters leading to full
disruption in both theories, the stronger tidal forces in relativity imply that the star is disrupted further from the black hole and
that the debris is therefore less tightly bound, leading to a smaller peak fallback accretion rate.  We also examine the capture
of tidal debris by the horizon and the relativistic pericenter precession of tidal debris, finding that black holes of $10^6$
solar masses and above generate tidal debris precessing by $10^\circ$ or more per orbit.
\end{abstract}
\maketitle

\section{Introduction} \label{S:intro}

Supermassive black holes (SBHs) with masses in the range from $10^{6}M_{\odot} \leq M_{\bullet} \leq 10^{10}M_{\odot}$
are found at the centers of most large galaxies \cite{1995ARA&A..33..581K,1998AJ....115.2285M}.  Observations of
quasars~\cite{1963Natur.197.1040S} and active galactic nuclei~(AGN) have aided our understanding of these massive
celestial objects, but these constitute only a small fraction~\cite{2003MNRAS.346.1055K} of all SBHs in the local universe,
making it difficult to take a fully representative census of galactic nuclei. Studying quiescent SBHs, visible via their
interactions with nearby stellar objects through the phenomena of tidal disruption events (TDEs)
\cite{1988Natur.333..523R}, would allow one to fill in this population gap, particularly for lower mass SBHs. Quiescent SBHs
also have dynamical effects on neighboring stars and gas, but these can only be observed locally where the influence
radius can be resolved.

TDEs occur when tidal forces due to a SBH's gravitational field overcome an orbiting star's self gravity, causing roughly half
of the resulting debris to become more tightly bound to the SBH \cite{1988Natur.333..523R}.  This bound debris quickly
forms a disk that is accreted by the SBH, releasing large amounts of energy in what is known as a tidal flare.  Flares can be
found in the optical
\cite{2011ApJ...741...73V,2012Natur.485..217G,2012aCenko,2014Arcavi,2014Chornock,2014Holoien,2015Vinko},
ultraviolet \cite{2004Stern,2006Gezari,2008Gezari,2009Gezari}, soft X-ray
\cite{1996Bade,1999Grupe,2002Donley,2008Esquej,2010Maksym,2012Saxton}, and hard X-ray
\cite{2011Bloom,2011Burrows,2012bCenko} bands. A recent survey~\cite{2002AJ....124.1308D} discovered many
candidate TDEs, showing rates consistent with those previously predicted by early models, resulting in an increased interest
in TDEs in recent years.

Analysis of TDEs has been conducted under both Newtonian gravity
\cite{1988Natur.333..523R,1999MNRAS.309..447M,2004ApJ...600..149W,2013CQGra..30x4005M,1978ApJ...226.1087C,1989Evans,2009MNRAS.392..332L} and general relativity
\cite{1985Luminet,1993Laguna,1997Diener,2003Ivanov,2006A&A...448..843I,2012PhRvD..85b4037K,Mike2nd,
2016MNRAS.458.4250S}.  In this paper, we
provide a unified, self-consistent treatment of several aspects of the tidal-disruption process that are sensitive to general
relativity.  We explore how relativity changes the boundaries of the loss cone in phase space that determines TDE rates and
the distribution of orbital elements for tidally disrupted stars.  We then examine how relativity affects the distributions of the
specific energy and angular momentum of the tidal debris and combine these results with recent hydrodynamical
simulations to predict the rate at which this debris falls back onto the SBH to fuel a tidal flare \cite{2013ApJ...767...25G}.

We restrict ourselves in this paper to the Schwarzschild spacetime \cite{1916SPAW.......189S} corresponding to
non-spinning SBHs.  The spherical symmetry of the Schwarzschild metric implies that only the magnitudes of the orbital
energy and angular momentum are needed to define the boundaries of the loss cone.  However, in the Kerr spacetime
\cite{1963PhRvL..11..237K} of spinning SBHs, the orientation of the tidally disrupted star's orbit with respect to the SBH's
equatorial plane affects the magnitude of the tidal stresses \cite{2006A&A...448..843I}. This creates a
higher-dimensional parameter space, the analysis of which is beyond the scope of this paper.  We will, however, provide
some comments on SBH spin when relevant throughout the paper and lay the foundation for future work that will focus on
incorporating spin dependence into our procedure.

The rest of this paper is structured as follows. Sec.~\ref{S:LCT} provides a brief overview of how loss-cone theory is used
to calculate TDE rates in Newtonian gravity.  Sec.~\ref{S:rel} details how relativity changes the boundaries of the loss
cone, allowing us to compare the TDE rates predicted in Newtonian gravity and general relativity.  We then focus on trying to
compare individual TDEs in the two gravitational theories, which introduces the somewhat subtle issue of which parameter
to hold constant in these comparisons.  We introduce a new procedure to map between Newtonian orbits and relativistic
geodesics in Sec.~\ref{S:map} that allows us to estimate how relativity would modify simulations performed in Newtonian
gravity.  In Sec.~\ref{S:prop}, we use this mapping and loss-cone theory to predict distributions of physical quantities like
the peak fallback accretion rate and amount of apsidal precession that affect observed TDE properties.  A summary of our
key results, their implications for TDE observations, and our plans to generalize to the Kerr spacetime are provided in
Sec.~\ref{S:disc}.

\section{Newtonian loss-cone theory} \label{S:LCT}

We begin by providing a brief summary of Newtonian loss-cone theory
\cite{1978ApJ...226.1087C,1999MNRAS.309..447M,2004ApJ...600..149W,2013CQGra..30x4005M}.  We define the tidal
radius as
\begin{equation} \label{E:rtdef}
	r_t \equiv \left( \frac{M_\bullet}{M_\star} \right)^{1/3}R_{\star}
\end{equation}
where $M_{\bullet}$ is the mass of the SBH, $M_{\star}$ is the mass of the tidally disrupted star, and $R_{\star}$ is the
stellar radius.  This definition provides an order-of-magnitude estimate of the distance from the SBH at which tidal disruption
occurs.  Stars on parabolic orbits with pericenters equal to $r_t$ will have orbital angular momentum
\begin{equation} \label{E:Ltdef}
	L_t \equiv (2GM_\bullet r_t)^{1/2}~.
\end{equation}
Any star on an orbit with angular momentum $L \lesssim L_{t}$ will become tidally disrupted once it reaches a distance
$r \sim r_{t}$ from the SBH. We can refine these estimates to account for stellar structure by introducing the parameter $\eta$
which measures the duration of the star's pericenter passage relative to the hydrodynamical time of the star
\cite{1977ApJ...213..183P,1989Evans}.  This parameter is a function of the polytropic index $\gamma$ of the star's equation
of state \cite{1939isss.book.....C}.  We may then define
\begin{subequations} \label{E:NewtLrd}
	\begin{align}
		r_d &\equiv \eta^{2/3}r_t~, \\
		L_d &\equiv \eta^{1/3}L_t~,
	\end{align}
\end{subequations}
the radial distance $r_{\rm d}$ and maximum orbital angular momentum $L_{\rm d}$ at which a star described by parameter
$\eta$ is tidally disrupted.  Throughout this paper, we use the value $\eta = 0.382$ consistent with the threshold for full tidal
disruption in Guillochon and Ramirez-Ruiz \cite{2013ApJ...767...25G} for a Solar-type star with $\gamma = 4/3$.

The TDE rate thus depends on the rate at which stars are driven onto such orbits, which are said to lie in the loss cone in
phase space because the velocity vector lies in a cone about the radial direction.  We assume that the stellar distributions in
galactic centers are described by isothermal spheres with density profile
\begin{equation} \label{E:rho}
\rho(r)=\frac{\sigma^{2}}{2\pi Gr^{2}}
\end{equation}
where $\sigma$ is the velocity dispersion given by the $M_{\bullet}-\sigma$ relation
\cite{2000ApJ...539L..13G,2000ApJ...539L...9F}.  This is a reasonable approximation for galaxies with cusps in the
surface brightness at their galactic centers \cite{2004ApJ...600..149W}.  We assume that loss-cone orbits are refilled by
two-body non-resonant relaxation \cite{1978ApJ...226.1087C,1999MNRAS.309..447M,2004ApJ...600..149W,2013CQGra..30x4005M}, in which case the differential flux $F$ of stars into the loss cone per unit dimensionless specific binding energy
$\varepsilon^{*} \equiv \varepsilon/\sigma^{2}$, following the approach and notation of Merritt \cite{2013degn.book.....M}, is
given by 
\begin{equation} \label{E:diffF}
F(\varepsilon^{*})=4\pi^{2}q(\varepsilon^{*})L^{2}_d \left[ \int_{0}^{y_d} f(\varepsilon^{*},y)dy \right]~.
\end{equation}
Here $q(\varepsilon^{*})$ is the orbital period divided by the time a star takes to diffuse across the loss cone,
$y \equiv L^2/qL_d^2$ is a dimensionless angular-momentum variable, $y_d \equiv 1/q$, and $f(\varepsilon^{*},y)$ is
the stellar phase-space density as a function of our dimensionless variables.  For the stellar density profile of
Eq.~(\ref{E:rho}), 
\begin{equation} \label{E:qISO}
q(\varepsilon^{*})=\frac{32\pi^{2}}{3\sqrt{2}} \ln \Lambda \left( \frac{M_{\star}}{M_{\bullet}} \right)
\frac{h(\varepsilon^{*})}{\psi^{*}(r_d)-\varepsilon^{*}} \left(\frac{r_d}{r_{h}} \right)^{-2} 
\end{equation}
where $\ln\Lambda$ is the Coulomb logarithm \cite{DynF1,DynF2,DynF3}, $\psi^{*}(r) \equiv -\Phi(r)/\sigma^2$ is the
negative dimensionless potential energy, $r_{h}=GM_{\bullet}/\sigma^{2}$ is the influence radius \cite{1972Peebles}, and
$h(\varepsilon^{*})$ (Eq.~(6.78) of Merritt \cite{2013degn.book.....M}) is a function derived from the isotropic distribution
function consistent with our isothermal stellar density profile $\rho(r)$.  We set $\Lambda=0.4M_{\bullet}/M_{\odot}$,
appropriate for stars whose velocity distribution is consistent with the virial theorem
\cite{1971ApJ...164..399S,2004ApJ...600..149W}.

The ratio $q$ helps provide intuition about how two-body relaxation feeds stars into SBHs.  Portions of phase space with
$q \ll 1$ are said to belong to the empty loss-cone (ELC) regime because diffusion is not efficient enough to repopulate
loss-cone orbits on their dynamical timescale.  Conversely, $q \gg 1$ corresponds to the full loss-cone (FLC) regime where
diffusion is effective enough to keep orbits in these parts of phase space filled.  SBHs with $M_{\bullet} \gtrsim 10^7\,
M_{\odot}$ have $q \ll 1$ for $\varepsilon^{*}>1$ and thus belong primarily to the ELC regime, implying lower TDE rates
despite their larger tidal radii \cite{2013CQGra..30x4005M}.  Approximately $30\%$ of TDEs occur in the FLC regime for the
galaxy sample considered in Stone and Metzger \cite{2016MNRAS.455..859S}, where $\sim 50\%$
result in full rather than partial disruptions.

The integrand in Eq.~(\ref{E:diffF}) for the isotropic distribution function consistent with the isothermal density profile of
Eq.~(\ref{E:rho}) is given by
\begin{equation} \label{E:DF}
f(y)=f(y_d) \left[ 1-\frac{2}{\sqrt{y_d}}\sum_{m=1}^{\infty}\frac{e^{-\gamma_m^{2}/4}}{\gamma_m}
\frac{J_0(\gamma_m\sqrt{y})}{J_1(\gamma_m\sqrt{y_d})} \right]
\end{equation}
where $J_0$ and $J_1$ are Bessel functions of the first kind and $\gamma_m$ is defined such that
$\gamma_m\sqrt{y_d}$ is the $m$-th zero of $J_0$ \cite{2013CQGra..30x4005M}.  The flux of stars into the loss cone is
then,
\begin{equation} \label{E:diffF2}
F(\varepsilon^{*})=4\pi^{2} L^2_d f(y_d) \xi(q)
\end{equation}
where $\xi(\varepsilon^{*})$ is defined to be $q$ times the integral in Eq.~(\ref{E:diffF}) and is well approximated by 
$\xi(q) \approx q/(q^{2}+q^{4})^{1/4}$.  The total TDE rate is obtained by integrating over all binding energies,
\begin{equation} \label{E:TDErate}
\dot N=\int F(\varepsilon^{*})\, d\varepsilon^{*},
\end{equation}
and depends implicitly on the SBH mass $M_\bullet$.

\section{Relativistic tidal forces} \label{S:rel}

In Newtonian gravity, the tidal force acting on a fluid element of a star is simply the gravitational force in a frame freely falling
with the star's center of mass.  In general relativity however, gravity affects particle motion through the curvature of
spacetime.  In this section, we review how tides arise in general relativity so that we can determine the value of the angular
momentum $L_d$ that sets the boundary of the loss cone in this gravitational theory.  We restrict our analysis in this paper to the
Schwarzschild metric \cite{1916SPAW.......189S} describing non-spinning SBHs, which in Boyer-Lindquist coordinates
\cite{1967JMP.....8..265B} is
\begin{align}
ds^{2} &= -\left( 1-\frac{2M_\bullet}{r} \right)dt^{2} + \frac{r^{2}}{r^{2}-2M_{\bullet}}dr^{2} \notag \\
& \quad + r^{2}(d \theta^{2}+\sin^{2}\theta d \phi^{2}).  \label{E:Schw}
\end{align}
Massive test particles experiencing no non-gravitational forces travel on timelike geodesics $x^\mu(\tau)$ of this metric, the
generalization of straight lines in flat space.  We can choose to parameterize these geodesics by the proper time $\tau$.
Particles traveling on these geodesics have conserved specific energy $E$ and specific angular momentum $L$ since the
Schwarzschild metric is both time independent and spherically symmetric.  Note that the specific energy
\begin{equation} \label{E:relE}
E \equiv -g_{\mu\nu}U^\mu \left( \frac{\partial}{\partial t} \right)^\nu = \left( 1-\frac{2M_\bullet}{r} \right) \frac{dt}{d\tau} \,,
\end{equation}
where
\begin{equation} \label{E:4V}
U^\mu = \frac{dx^\mu}{d\tau} = \left( \frac{dt}{d\tau},  \frac{dr}{d\tau},\frac{d\theta}{d\tau},\frac{d\phi}{d\tau} \right)
\end{equation}
is the 4-velocity, asymptotes to unity for particles at rest far from the SBH since it contains the rest-mass energy.  We
approximate $E = 1$ for the orbits of tidally disrupted stars because the velocity dispersion $\sigma$ of their host galaxies
is much less than the speed of light.  This 4-velocity satisfies the geodesic equation
\begin{equation} \label{E:geo}
\frac{dU^\mu}{d\tau} + \Gamma^\mu_{\nu\alpha} U^\nu U^\alpha = 0
\end{equation}
where $\Gamma^{\mu}_{\nu \alpha}$ are the Christoffel symbols for the Schwarzschild metric.  Note the resemblance to
Newton's second law $d\mathbf{v}/dt = 0$ for force-free motion for vanishing Christoffel symbols as can be chosen for flat
space.

In general relativity, tidal forces arise because of the tendency of parallel geodesics to deviate from each other in the
presence of spacetime curvature.  If two neighboring particles with 4-velocity $U^\mu$ are separated by an infinitesimal
spacelike deviation 4-vector $X^\beta$, this deviation will evolve with proper time $\tau$ according to the geodesic deviation
equation
\begin{align} 
\frac{d^2X^\beta}{d\tau^2} &= U^\mu \nabla_\mu (U^\alpha \nabla_\alpha X^\beta) = -R^\beta_{~\mu\alpha\nu} U^\mu
X^\alpha U^\nu \notag \\
&= -C^\beta_{~\alpha} X^\alpha \label{E:geoD}
\end{align}
where $\nabla_\mu$ are covariant derivatives, $d/d\tau \equiv U^\alpha \nabla_\alpha$ is the derivative with respect to the
proper time, $R^{\beta}_{~\mu \alpha \nu}$ is the Riemann curvature tensor, and
\begin{equation} \label{E:tidaltensor}
C^{\beta}_{~\alpha} \equiv R^{\beta}_{~\mu \alpha \nu}U^{\mu}U^{\nu}
\end{equation}
is the tidal tensor.  In Newtonian gravity, a fluid element is tidally stripped from the surface of a star when the tidal force
exerted by the SBH exceeds the force exerted on this fluid element by that star's self gravity.  If the star is non-degenerate
like the Solar-type stars considered in this paper, we can approximate its self gravity as a Newtonian force even if general
relativity is needed to describe the geodesics of the SBH.  In this approximation, the fluid element is tidally stripped when the
acceleration due to this self gravity is exceeded by the proper acceleration $d^2X/d\tau^2$ given by the geodesic deviation
equation (\ref{E:geoD}) \cite{1983grg1.conf..438L}.

We can solve this equation more easily by transforming from the global Boyer-Lindquist coordinates given by
Eq.~(\ref{E:Schw}) to local Fermi normal coordinates ($\tau, X^{(i)}$) valid in a neighborhood of spacetime about the center
of mass of the star at an arbitrarily chosen proper time $\tau = 0$ \cite{FermiNormcoords}.  These coordinates define an
orthonormal tetrad: the star's 4-velocity provides the timelike 4-vector $\lambda^\mu_{~(0)}$, and three spacelike 4-vectors
$\lambda^\mu_{~(i)}$ are chosen that are parallel transported along the geodesic.  Spatial indices that run from 1 to 3 are
denoted by Latin indices, unlike Greek indices that run from 0 to 3.  All of the tensors appearing in the geodesic deviation
equation can be projected into this basis:
\begin{subequations} \label{E:FNC}
\begin{align}
U^\mu &= \lambda^\mu_{~(0)} \\
X^\alpha &= X^{(i)} \lambda^\alpha_{~(i)} \\
R^{\beta}_{~\mu \alpha \nu} &= R^{(\gamma)}_{\quad(\delta)(\kappa)(\xi)} \lambda^\beta_{~(\gamma)}
\lambda_\mu^{~(\delta)} \lambda_\alpha^{~(\kappa)} \lambda_\nu^{~(\xi)} \\
C^{\beta}_{~\alpha} &= C^{(i)}_{\quad(j)} \lambda^\beta_{~(i)} \lambda_\alpha^{~(j)}
= R^{(i)}_{\quad(0)(j)(0)} \lambda^\beta_{~(i)} \lambda_\alpha^{~(j)}.
\end{align}
\end{subequations}
The Boyer-Lindquist indices are raised and lowered by the Schwarzschild metric (\ref{E:Schw}) while the Fermi normal
coordinate indices in brackets are raised and lowered by the Lorentz metric $\eta_{(\mu)(\nu)}$.  Note that the symmetry of
the Riemann tensor implies that the tidal tensor is symmetric and only has spatial components in Fermi normal coordinates.
Inserting Eq.~(\ref{E:FNC}) into (\ref{E:geoD}) yields the geodesic deviation equation in Fermi normal coordinates
\begin{equation} \label{E:deoDFNC} 
\frac{d^2X^{(i)}}{d\tau^2} = -C^{(i)}_{\quad(j)} X^{(j)}.
\end{equation}
Because $C^{(i)}_{\quad(j)}$ is a real, symmetric $3\times 3$ tensor, it has 3 real eigenvalues which in Boyer-Lindquist
coordinates are $M_\bullet/r^3$, $(1+3L^2/r^2)M_\bullet/r^3$, and $-2(1+3L^2/2r^2)M_\bullet/r^3$.  The negative sign in
Eq.~(\ref{E:deoDFNC}) implies that the eigenvectors associated with the positive eigenvalues correspond to directions
along which the star is compressed, while the eigenvector associated with the negative eigenvalue corresponds to the
direction in which the star is stretched.  Equating the magnitude of this negative eigenvalue times the stellar radius
$R_\star$ with the acceleration $2M_\star/(\eta R_\star)^2$ due to the star's self gravity yields
\begin{equation} \label{E:tidalC}
\left(1+ \frac{3L^2}{2r^2} \right) \frac{2M_\bullet R_\star}{r^3} = \frac{2M_\star}{(\eta R_\star)^2}.
\end{equation}
Combining this result with the relation
\begin{equation} \label{E:rpL}
L^2 = \frac{2M_\bullet r^2}{r - 2M_\bullet},
\end{equation}
between the orbital angular momentum $L$ and Boyer-Lindquist coordinate $r$ at pericenter provides two equations
that can be solved for the the relativistic tidal radius $r_d$ and angular-momentum threshold $L_d$ for tidal disruption.
Note that in the limit $r \gg 2M_\bullet$ these reduce to the Newtonian expressions of Eqs.~(\ref{E:Ltdef}) and
(\ref{E:NewtLrd}).

\begin{figure}[t!]
\centering
\includegraphics[width=3.5in,height=3.25in]{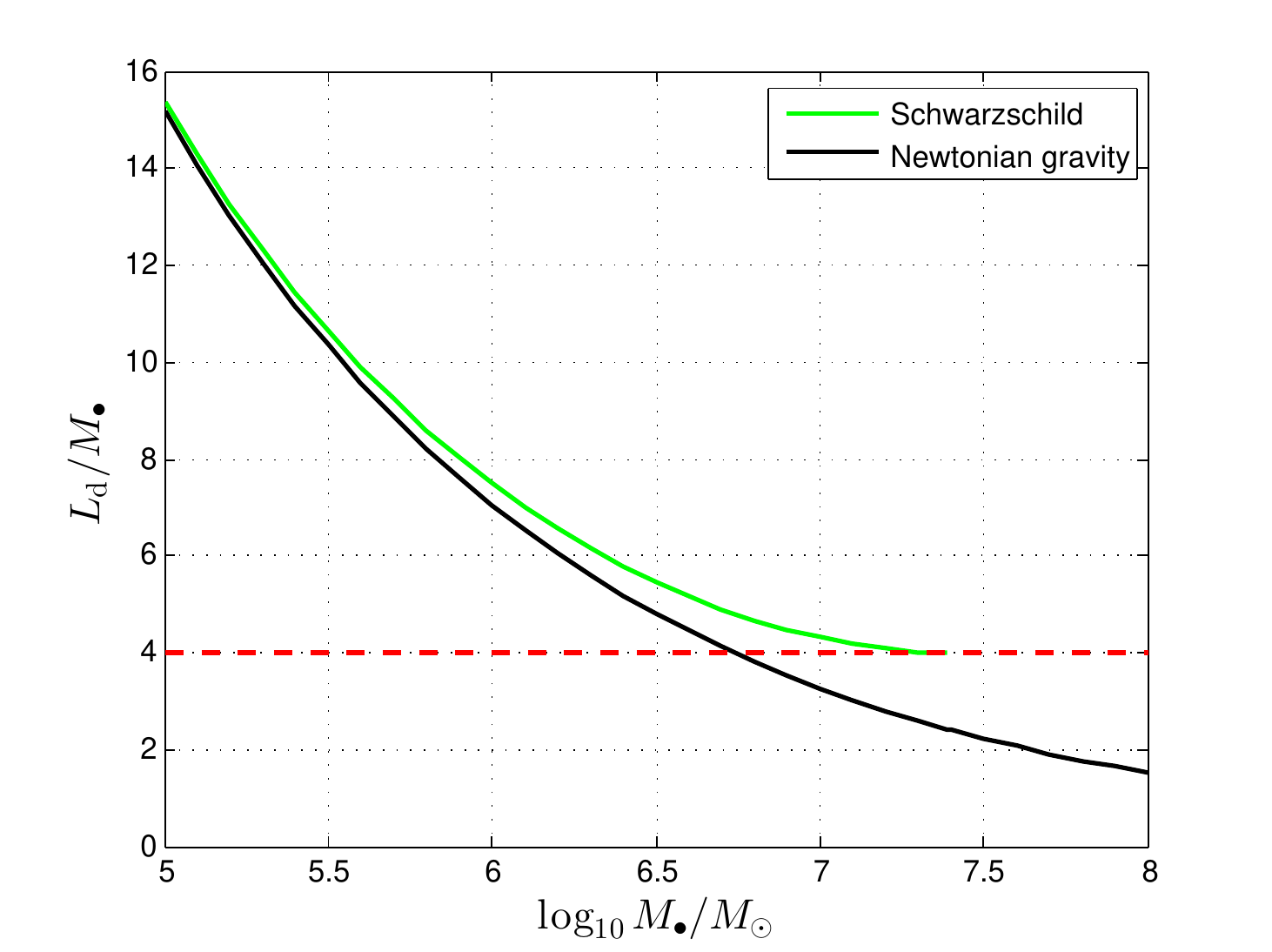}
\caption{The angular-momentum threshold $L_d$ for tidal disruption as a function of SBH mass $M_\bullet$ in both
Newtonian gravity (solid black) and general relativity (solid green).  As the mass of the SBH increases, $L_d$ falls less
steeply in relativity than in Newtonian gravity because of the stronger tidal forces.  The horizontal red dashed line at
$L_{\rm cap} = 4M_\bullet$ indicates that direct capture by the event horizon occurs below this value, implying that SBHs
with $M_\bullet > M_{\rm max} \simeq 10^{7.39} M_\odot$ cannot fully disrupt Solar-type stars.}  \label{F:LdM}
\end{figure}

\begin{figure}[t!]
\centering
\includegraphics[width=0.52\textwidth]{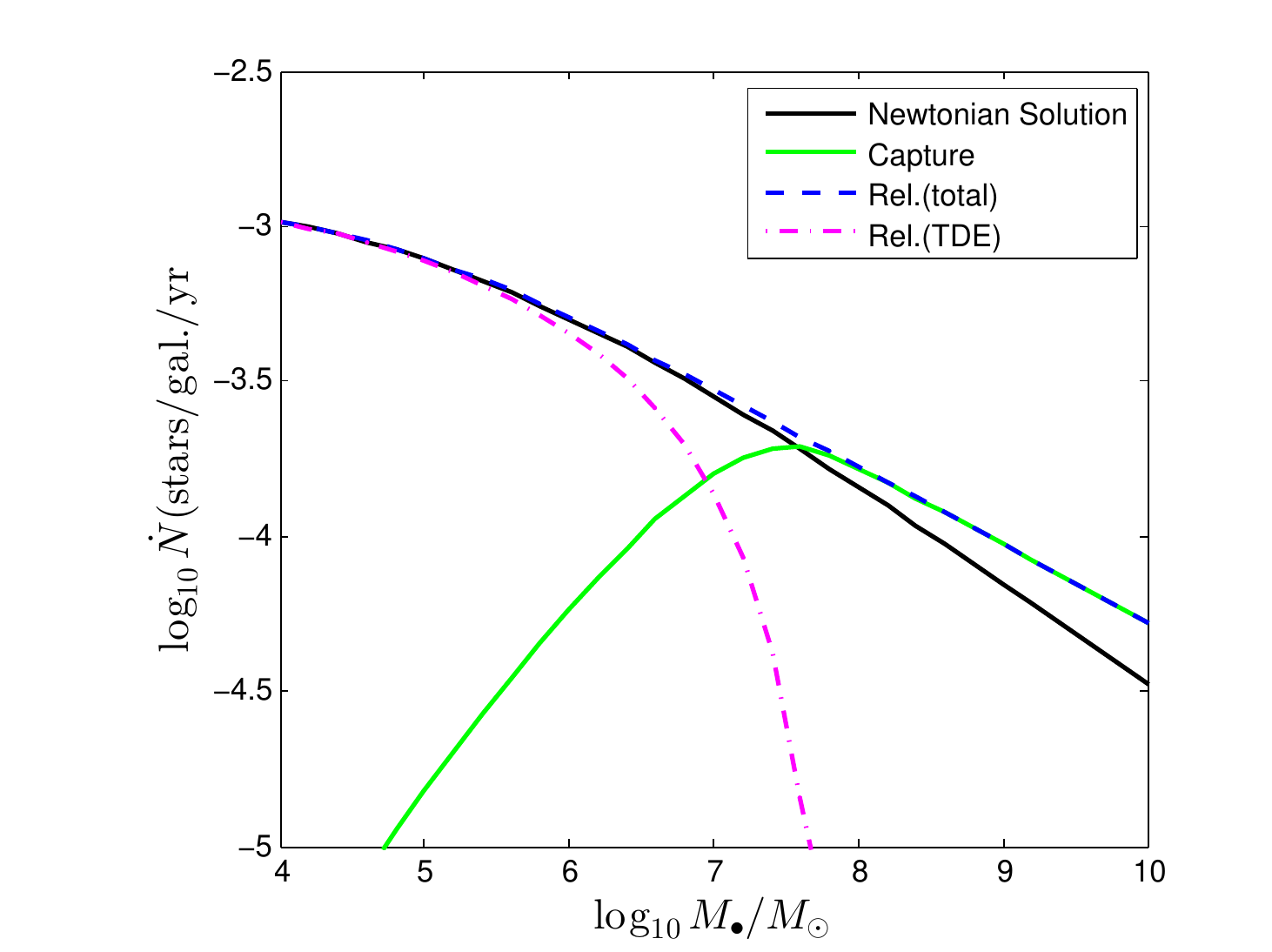}
\caption{TDE rates in Newtonian gravity and general relativity as functions of SBH mass $M_\bullet$.  The solid black curve
shows the TDE rate in Newtonian gravity.  In relativity, observable TDEs only result when the tides are strong enough to
disrupt the star but the angular momentum is above the threshold $L_{\rm cap}$ for direct capture by the event horizon.  The
rate of such observable TDEs is given by the dot-dashed magenta curve, while the capture rate is shown by the solid green
curve (the dashed blue curve shows the total of these two rates).  Stronger tides in relativity slightly increase this total rate,
but capture reduces the observed TDE rate in relativity below that of Newtonian gravity.} 
\label{F:TDErates}
\end{figure}

Fig.~\ref{F:LdM} shows how the angular-momentum threshold $L_d$ depends on SBH mass $M_\bullet$ in both Newtonian
gravity and general relativity.  All orbits with angular momentum $L$ below this threshold are considered to be inside the
loss cone.  The Newtonian tidal radius $r_t \propto M_\bullet^{1/3}$ according to Eq.~(\ref{E:rtdef}), while the gravitational
radius $r_g \equiv GM_\bullet/c^2$ scales linearly with SBH mass.  This implies that relativistic effects become negligible for
small SBH masses and the two curves in Fig.~\ref{F:LdM} converge towards the left edge of the plot.  The fact that the term
in brackets on the left-hand side of Eq.~(\ref{E:tidalC}) is greater than unity implies that tidal forces are stronger in general
relativity than in Newtonian gravity for orbits with the same angular momentum.  This explains why the relativistic curve in
Fig.~\ref{F:LdM} is above the Newtonian curve; stronger tides allow the SBH to disrupt stars at larger angular momentum
$L_d$ in relativity.

Relativity also introduces the phenomena of stellar capture; stars on orbits with angular momentum less than
$L_{\rm cap} = 4M_\bullet$ will plunge directly into the event horizon leaving no debris to emit photons and generate an
observable TDE.  This angular-momentum threshold for direct capture, shown by the horizontal red dashed line in
Fig.~\ref{F:LdM}, grows linearly with SBH mass and eventually exceeds the threshold $L_d$ for tidal disruption at
$M_\bullet = M_{\rm max} \simeq 10^{7.39} M_\odot$.  This prevents observable TDEs by SBHs with masses above this
value as shown by the intersection of the solid green and dashed red curves in Fig.~\ref{F:LdM}.

Fig.~\ref{F:TDErates} shows the rates for observable TDEs and direct capture predicted by Eq.~(\ref{E:diffF}) when two-body
relaxation is responsible for refilling the loss cone.  Relativity modifies the Newtonian prediction in two ways: (1) it changes
the thresholds $r_d$ and $L_d$ in Eqs.~(\ref{E:qISO}) and (\ref{E:DF}) to the maximum of the two thresholds set by tidal
disruption and direct capture, and (2) the limits of the integral in Eq.~(\ref{E:diffF}) must be modified.  If
$L_{\rm cap} < L_d$ ($M_\bullet < M_{\rm max}$), the integral must be decomposed into two parts: an integral from 0 to
$y_{\rm cap}$ giving the capture rate and an integral from $y_{\rm cap}$ to $y_d$ giving the observable TDE rate.  For
$L_{\rm cap} \geq L_d$ ($M_\bullet \geq M_{\rm max}$), there are no observable TDEs and a single integral from 0 to
$y_{\rm cap}$ gives the capture rate.

\section{Comparing tidal disruption in Newtonian gravity and general relativity} \label{S:map}

Having introduced tidal disruption in Newtonian gravity and general relativity in Secs.~\ref{S:LCT} and \ref{S:rel}, we now
seek to compare the predictions of the two theories.  This point is more subtle than it might initially appear, as there is no
unique way to map Keplerian orbits in Newtonian gravity to geodesics of the Schwarzschild metric in general relativity.  In
the limit that the pericenter $r \to \infty$, it is natural to map a geodesic to its identical Keplerian counterpart, but this limiting
behavior is insufficient to fully specify the mapping.  The appropriate mapping to use depends on the nature of the problem
one is trying to solve.  We consider in this section three distinct mappings that all possess the desired behavior in the
Newtonian limit; these three mappings identify:
\begin{itemize}

\item[(1)] orbits with equal pericenter coordinates $r$,

\item[(2)] orbits with equal angular momenta $L$,

\item[(3)] orbits on which a star experiences equal tidal forces at pericenter.

\end{itemize}

Mapping (1) is perhaps the most obvious choice and was used for the corrections to the orbital constants provided in
Kesden \cite{Mike2nd}.  However, according to the Schwarzschild metric in Boyer-Lindquist coordinates given in
Eq.~(\ref{E:Schw}), the physical significance of this mapping is that it identifies parabolic orbits such that the circular orbits in
the two gravitational theories with the same pericenter $r$ would have the same circumference $2\pi r$.  It is unclear why
this choice of mapping would be particularly useful for analyzing the tidal disruption of stars on non-circular orbits.

Mapping (2) identifies orbits in the two theories with the same values of the gauge-invariant orbital angular momentum, 
defined for the Schwarzschild metric as
\begin{equation} \label{E:Ldef} 
L \equiv g_{\mu\nu}U^\mu \left( \frac{\partial}{\partial\phi} \right)^\nu = r^2 \frac{d\phi}{d\tau}
\end{equation}
for equatorial orbits ($\theta = \pi/2$).  This mapping seems like a useful choice, as TDE properties do depend on the
orbital angular momentum as seen in the recent simulations of Guillochon and Ramirez-Ruiz \cite{2013ApJ...767...25G}.
However, tides are stronger on Schwarzschild geodesics than on Keplerian orbits with the same value of $L$ as
established in Sec.~\ref{S:rel}, so it seems unlikely that TDEs on orbits in the two theories identified in this manner would
have the same properties.

We conjecture that TDEs resulting from stars initially on orbits identified by mapping (3) will have similar properties because
the stars were subjected to similar tidal forces.  This conjecture is supported by the "freezing" model of Lodato, King, and
Pringle \cite{2009MNRAS.392..332L} which proposed that the orbital energy distribution of tidal debris is frozen in
following an instantaneous tidal disruption of an unperturbed star.  This model was used with reasonable success to
describe the light curve of the TDE PS1-10jh \cite{2012Natur.485..217G}.  We will use this freezing assumption later in
Sec.~\ref{S:prop} to determine the appropriate relativistic correction to the energy distribution, but we will apply the
correction at the disruption radius $r_d$ rather than pericenter.  This is consistent with the analytic model of Stone, Sari, and
Loeb \cite{2013MNRAS.435.1809S} and some hydrodynamical simulations of TDEs
\cite{2013MNRAS.434..909H,2013ApJ...767...25G} which showed that the spread in debris energy was largely independent
of the orbital angular momentum $L$ for full disrupted stars.

To implement mapping (3), we must find the angular momentum $L_N$ of the orbit in Newtonian gravity that has the same
peak tidal force as that experienced by a star on a Schwarzschild geodesic with angular momentum $L$ and pericenter $r$.
We accomplish this by setting the magnitude of the negative eigenvalue of the tidal tensor $C^{(i)}_{\quad(j)}$ equal to its
Newtonian limit for an orbit with angular momentum $L_N$:
\begin{equation} \label{E:LN}
\left( 1+\frac{3L^2}{2r^2} \right) \frac{1}{r^3}= \left( \frac{2M_\bullet}{L_N^2} \right)^3.
\end{equation}
This equation, combined with Eq.~(\ref{E:rpL}) relating the angular momentum $L$ to the pericenter $r$ in Boyer-Lindquist
coordinates, can be solved to determine $L_N(L)$ for mapping (3).

\begin{figure}[t!]
\centering
\includegraphics[width=0.53\textwidth]{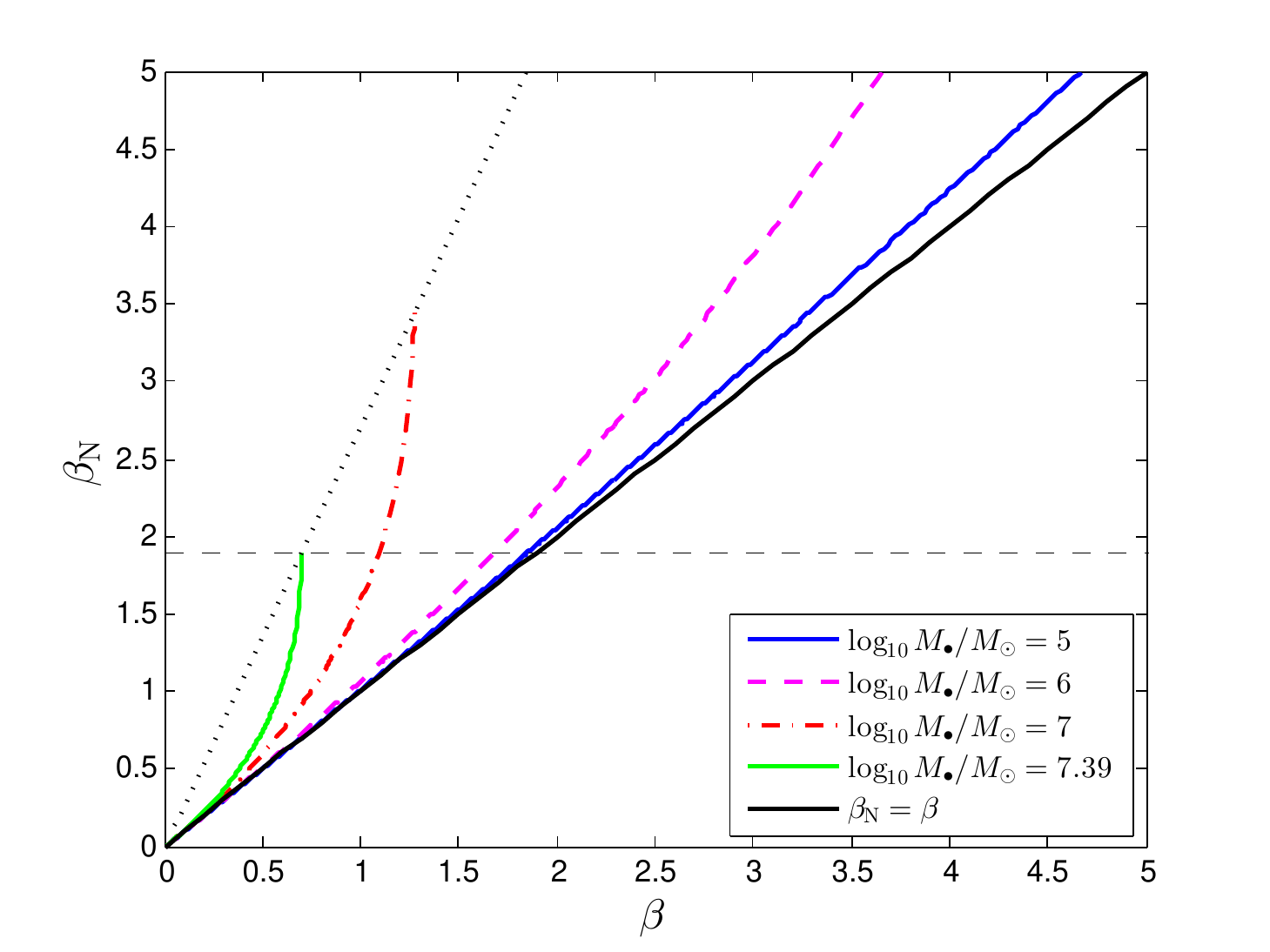}
\caption{The mapping between Schwarzschild geodesics with penetration factor $\beta$ and orbits in Newtonian gravity
with penetration factor $\beta_N$ on which stars experience the same tidal forces at pericenter.  The solid blue, dashed
magenta, dot-dashed red, and solid green curves show this mapping for SBHs with masses $M_\bullet/M_\odot$ of $10^5$,
$10^6$,  $10^7$, and $10^{7.39}$ respectively.  The solid black diagonal $\beta_N = \beta$ shows the limit of these curves
as $M_\bullet \to 0$, while the dotted black curve shows the relation $\beta_N(\beta_{\rm cap})$ beyond which these
mappings are undefined.  The horizontal black dashed line $\beta_{N,d} = 1.9$ corresponds to full disruption in the
hydrodynamical simulations of Guillochon and Ramirez-Ruiz \cite{2013ApJ...767...25G}.}
\label{F:betaN}
\end{figure}

Instead of using the angular momentum $L$ to identify orbits, the TDE literature often uses the penetration factor
\begin{equation} \label{E:beta}
\beta \equiv \frac{L_t^2}{L^2},
\end{equation}
where $L_t$ is defined in Eq.~(\ref{E:Ltdef}).  This choice is convenient because stars on orbits with
$\beta \geq \beta_d \equiv L_t^2/L_d^2$ are tidally disrupted, where $\beta_d$ is of order unity.  Our mapping $L_N(L)$
implies an equivalent mapping $\beta_N(\beta)$, where
\begin{equation} \label{E:bNb}
\beta_N \equiv \frac{L_t^2}{L_N^2} = \left( \frac{r + M_\bullet}{r - 2M_\bullet} \right)^{1/3} \frac{r_t}{r}
\end{equation}
and $r$ is an implicit function of $L$ and hence $\beta$ through Eq.~(\ref{E:rpL}). We show this mapping for several values of
the SBH mass $M_\bullet$ in Fig.~\ref{F:betaN}.  These mappings are undefined for
$\beta > \beta_{\rm cap} \equiv L_t^2/L_{\rm cap}^2$ since such geodesics plunge directly into the event horizon and
therefore do not have pericenters at which one can calculate the tidal force.  The mappings $\beta_N(\beta)$ are above the
diagonal and have positive curvature because tides are stronger in general relativity, requiring Newtonian orbits to have
deeper penetration factors $\beta_N$ to match the tides on Schwarzschild geodesics at pericenter.  The mapping for SBH
mass $M_{\rm max} \simeq 10^{7.39} M_\odot$ terminates at $\beta_N(\beta_{\rm cap}) = \beta_{N,d} = 1.9$, precisely the
threshold for the full disruption of Solar-type stars \cite{2013ApJ...767...25G} seen in Fig.~\ref{F:LdM} for
$L_d = L_{\rm cap}$.

\begin{figure}[t!]
\centering
\includegraphics[width=0.53\textwidth]{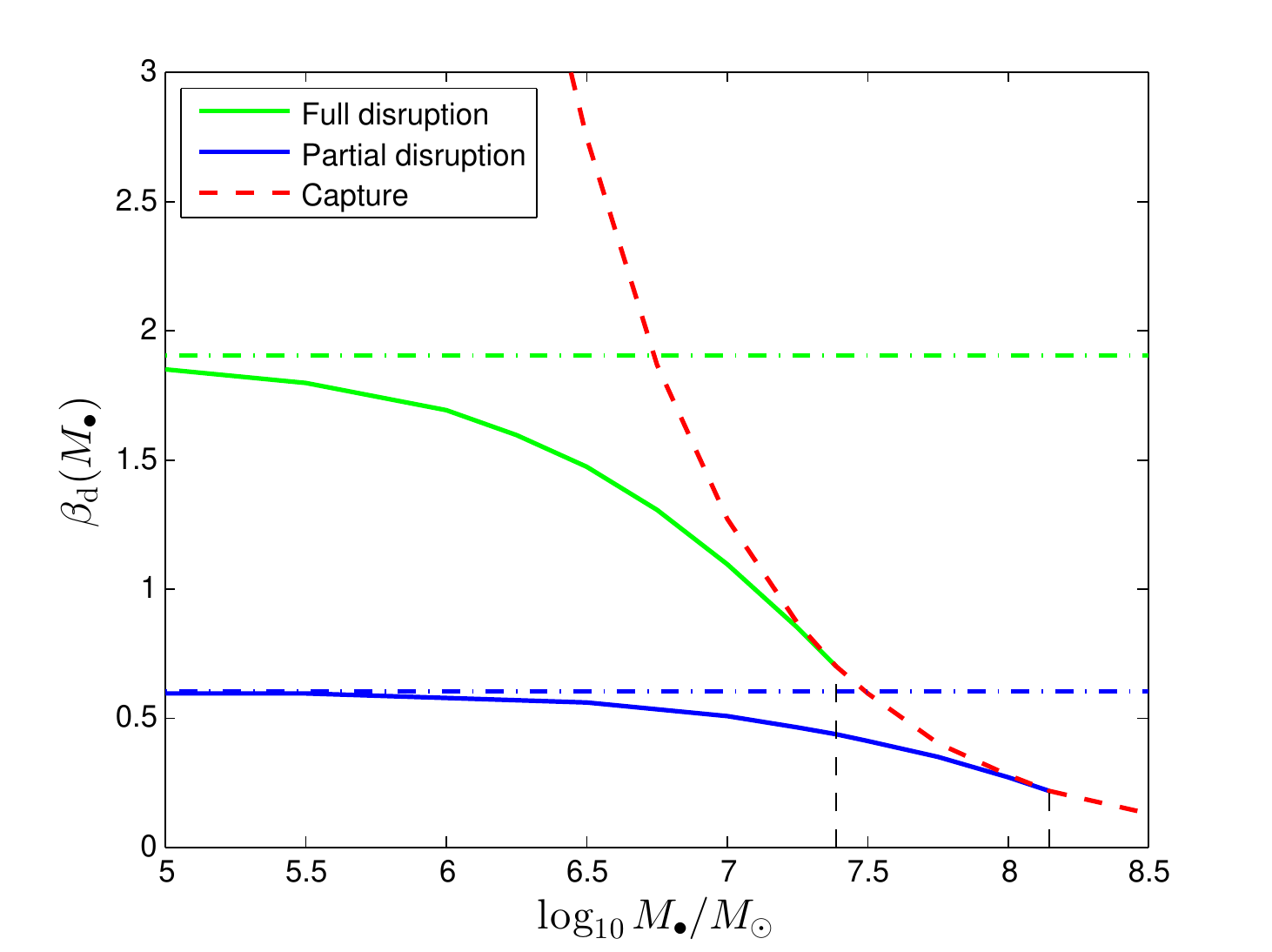}
\caption{The minimum penetration factors $\beta_d$ and $\beta_{pd}$ for full and partial tidal disruption in general relativity
as functions of SBH mass $M_\bullet$.  The horizontal dot-dashed green and blue lines show the thresholds
$\beta_{N,d} = 1.9$ and $\beta_{N,pd} = 0.6$ for full and partial disruption in Newtonian gravity \cite{2013ApJ...767...25G}.
The solid green and blue curves show the values of $\beta_d$ for Schwarzschild geodesics on which stars experience the
same tidal forces at pericenter.  The intersections between these curves and the dashed red curve
$\beta_{\rm cap}(M_\bullet)$, marked by the vertical dashed lines, occur at the maximum SBH masses
$M_{\rm max, FD} \simeq 10^{7.39}M_\odot$ and $M_{\rm max, PD} \simeq 10^{8.15}M_\odot$ capable of full and partial
tidal disruption.} \label{F:betaDM}
\end{figure}

To further illustrate how relativity affects the minimum penetration factor $\beta_d$ for tidal disruption, we show its
dependence on SBH mass $M_\bullet$ in Fig.~\ref{F:betaDM} for both full and partial tidal disruptions.  Newtonian
hydrodynamical simulations \cite{2013ApJ...767...25G} indicate that Solar-type stars are fully disrupted for
$\beta_N > \beta_{N,d} = 1.9$, while partial disruptions occur for $\beta_N > \beta_{N,pd} = 0.6$.  As $M_\bullet$ increases
and $\beta_d$ approaches $\beta_{\rm cap}$, the $\beta_d$ curves fall increasingly below these Newtonian thresholds
because of the stronger tides in general relativity. The values of $M_\bullet$ at which these curves intersect the
$\beta_{\rm cap}$ curve are the maximum masses $M_{\rm max, FD} \simeq 10^{7.39}M_\odot$ and
$M_{\rm max, PD} \simeq 10^{8.15}M_\odot$ capable of full and partial tidal disruption.  These limits are consistent with
recent work suggesting that observable TDEs by SBHs with masses above $10^8 M_\odot$ result exclusively from giant
stars \cite{GiantTDEs}.  Determinations of $M_{\rm max}$ that failed to account for the stronger relativistic tides, effectively
using mapping (2) $\beta_d = \beta_{N,d}$ shown by the horizontal lines in Fig.~\ref{F:betaDM}, would underestimate
$M_{\rm max}$ by a factor of $\sim 4.5$ for both full and partial disruptions.  Future work will explore how SBH spin can push
$M_{\rm max}$ to even higher values for stars on prograde orbits.

\section{Distributions of relativistic TDE properties} \label{S:prop}

\begin{figure*}[t]
\subfloat{\includegraphics[width=3.57in,height=3.3in]{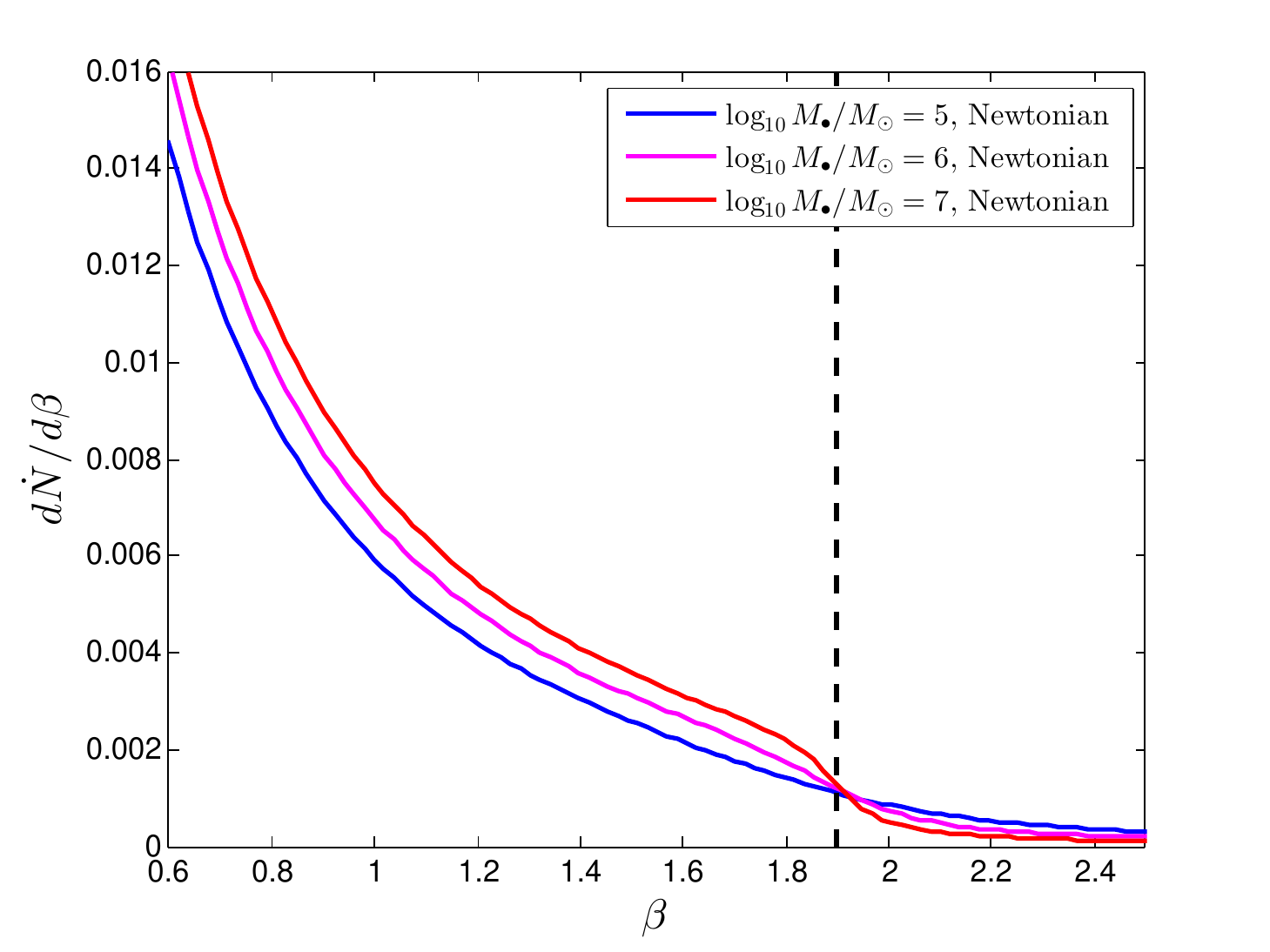}}
\subfloat{\includegraphics[width=3.7in,height=3.3in]{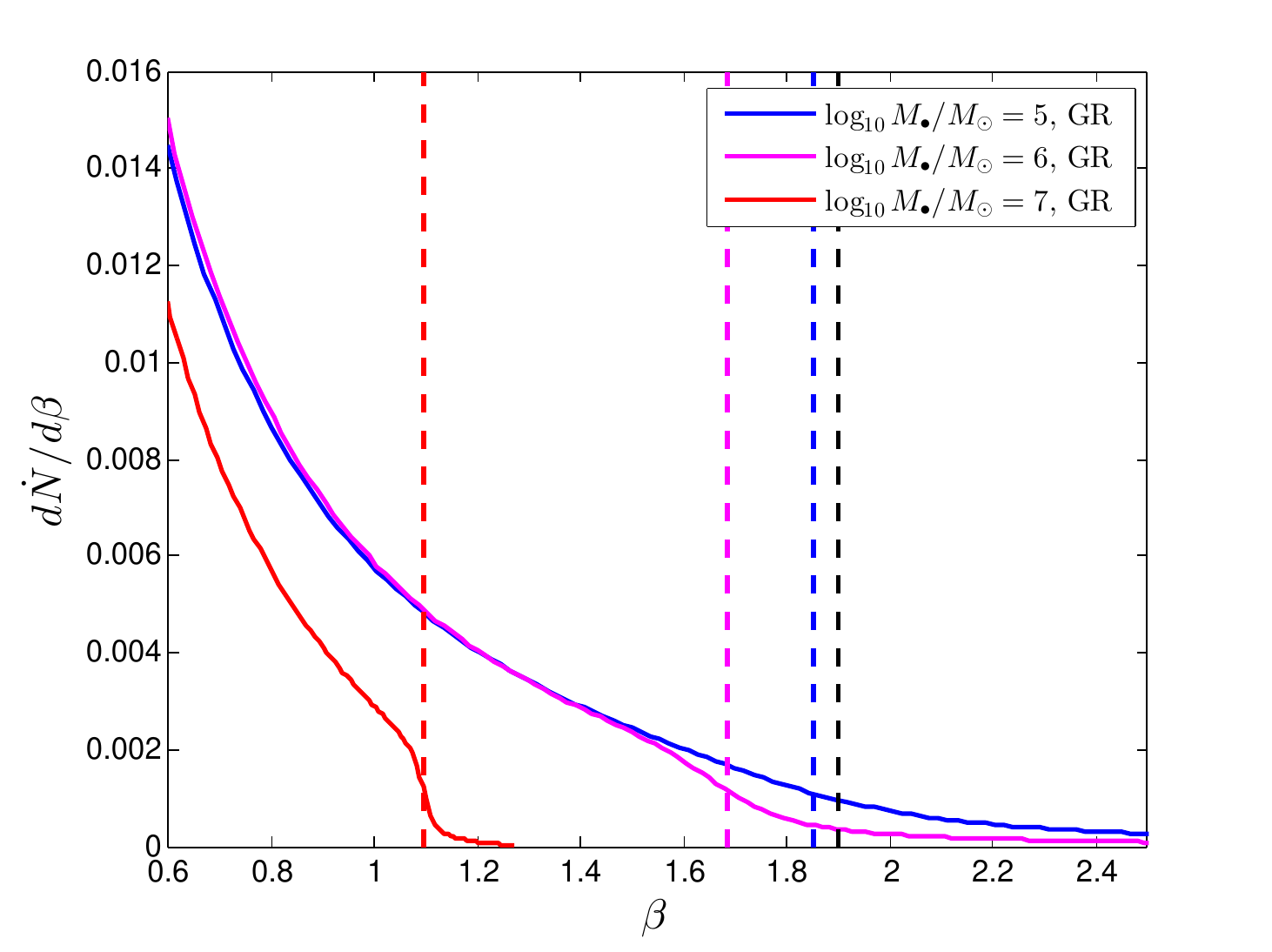}}
\caption{\label{fig:5} The differential TDE rate $d\dot N/d\beta$ per unit penetration factor for various SBH masses under
Newtonian gravity (left) and relativity (right).  The dashed black line in both panels shows the Newtonian threshold
$\beta_{N,d} = 1.9$ for full disruption, while the blue, magenta, and red dashed lines in the right panel show the thresholds
$\beta_d$ in relativity defined through the the mapping $\beta_N(\beta_d) = \beta_{N,d}$ between Newtonian orbits
and Schwarzschild geodesics with the same tidal forces at pericenter.} \label{F:dNdbeta}
\end{figure*}

In this section, we use concepts developed earlier in the paper to predict distributions of several quantities that affect the
observed properties of TDEs.  Guillochon and Ramirez-Ruiz \cite{2013ApJ...767...25G} performed Newtonian
hydrodynamical simulations to determine the peak accretion rate $\dot{M}_{\rm peak}$ at which tidal debris falls back on the
SBH following a TDE and the delay $t_{\rm peak}$ between the tidal disruption and the time at which this peak accretion
occurs.  They then provided analytic fits to these quantities as functions of the Newtonian penetration factor $\beta_N$. 
Under the approximation that these quantities depend only on the peak tidal forces along each orbit, we use the inverse of
the mapping $\beta_N(\beta)$ shown in Fig.~\ref{F:betaN} to derive fits to $\dot{M}_{\rm peak}$ and $t_{\rm peak}$ as
functions of the relativistic penetration factor $\beta$ in the Schwarzschild spacetime.  We then use the loss-cone theory
discussed in Sec.~\ref{S:LCT}, modified by the relativistic corrections to the boundaries of the loss cone shown in
Fig.~\ref{F:LdM}, to predict the distributions of these quantities for different SBH masses.

In Subsec.~\ref{SS:dbeta}, we determine the differential TDE rate $d\dot{N}/d\beta$ per unit relativistic penetration factor
$\beta$.  In Subsec.~\ref{SS:dpeak}, we use this rate and our corrected fits $\dot{M}_{\rm peak}(\beta)$ and
$t_{\rm peak}(\beta)$ to derive initial estimates for the differential TDE rates $d\dot{N}/d\dot{M}_{\rm peak}$ and
$d\dot{N}/dt_{\rm peak}$ per unit peak fallback accretion rate and time delay respectively.  These initial estimates do not
account for an additional relativistic effect discussed in Kesden \cite{Mike2nd}, that the gradient of the potential well that
determines the width of debris energy distribution differs in general relativity and Newtonian gravity.  We review this result
and then derive revised estimates for $d\dot{N}/d\dot{M}_{\rm peak}$ and $d\dot{N}/dt_{\rm peak}$ in Subsec.~\ref{SS:relE}.
The stellar debris produced following tidal disruption will also have a distribution of orbital angular momentum, but this
often receives less attention in Newtonian gravity where the orbital periods that determine the fallback accretion rate depend
on energy but not angular momentum.  However, the distribution of orbital angular momentum can be very important when
considering highly relativistic TDEs for which much of the debris can lose enough specific angular momentum to fall below
the threshold for capture $L_{\rm cap}$ even if the initial star was above this threshold.  We discuss this regime in
Subsec.~\ref{SS:relL} and show that this effect sharply limits the relativistic correction to the peak fallback accretion rate.
Finally, in Subsec.~\ref{SS:omega} we use our relativistically corrected loss-cone theory to determine the differential TDE
rate $d\dot{N}/d\Delta\omega$ per unit relativistic shift in argument of pericenter, an important effect responsible for tidal
stream crossings and the prompt circularization of debris
\cite{2013MNRAS.434..909H,2013ApJ...775L...9D,2014ApJ...783...23G,2015ApJ...809..166G,2015ApJ...804...85S}.

\subsection{Distribution of penetration factor $\beta$} \label{SS:dbeta}

We begin by calculating the differential TDE rate $d\dot{N}/d\beta$ in Newtonian gravity.  The angular-momentum variable
$y$ can be expressed in terms of the penetration factor $\beta$ as
\begin{equation} \label{E:ybeta}
y=\frac{L_t^2}{qL_d^2\beta}~.
\end{equation}
This relation allows us to differentiate Eq.~(\ref{E:diffF}) with respect to $\beta$, which when combined with the distribution
function in Eq.~(\ref{E:DF}) yields
\begin{align} \label{E:dFdbeta}
\frac{dF}{d\beta} &= \left( \frac{2\pi L_t}{\beta} \right)^2 f(y_d) \times \notag \\
& \quad \left[ 1 - 2\sqrt{q} \sum_{m=1}^{\infty} \frac{e^{-\gamma_m^{2}/4}}{\gamma_m}
\frac{J_{0}(\gamma_m L_t/L_d\sqrt{q\beta})}{J_{1}(\gamma_m/\sqrt{q})} \right]\, .
\end{align}
Integrating this expression with respect to the dimensionless specific binding energy $\varepsilon^\ast$ gives us the
Newtonian differential TDE rate $d\dot{N}/d\beta$ shown in the left panel of Fig.~\ref{F:dNdbeta} above.  The total TDE rate
is the area under this curve for $\beta_N > \beta_{N,d}$ which decreases with increasing SBH mass $M_\bullet$ consistent
with the TDE rates seen in Fig.~\ref{F:TDErates}.  The blue curve in the left panel of Fig.~\ref{F:dNdbeta} extends smoothly
beyond $\beta_{N,d}$ because much of the phase space beyond this value is still in the full loss-cone regime $q \gg 1$ for
which $d\dot N/d\beta \propto \beta^{-1}$ \cite{2016MNRAS.455..859S}.  However, as $M_\bullet$ increases towards
$10^7 M_\odot$, the differential TDE rate becomes exponentially suppressed for $\beta_N > \beta_{N,d}$ reflecting that
much of the phase space beyond this point now lies in the empty loss-cone regime \cite{2016MNRAS.455..859S}.

Relativity introduces two changes into the calculation of the differential TDE rate $d\dot{N}/d\beta$: (1) the stronger tides
increase the numerical value of $L_d$ from the Newtonian result given by the black curve in Fig.~\ref{F:LdM} to the
relativistic result given by the green curve in this figure, and (2) capture by the event horizon cause the differential rate to
fall discontinuously to zero for $\beta \geq \beta_{\rm cap}$.  As most stars diffusing into the loss cone have apocenters near
the influence radius $r_h = GM_\bullet/\sigma^2 \gg r_g = GM_\bullet/c^2$ \cite{1976MNRAS.176..633F}, the Newtonian
treatment of stellar diffusion encapsulated in the ratio $q$ remains an accurate approximation.  Both of the changes listed
above leave signatures in the relativistic differential TDE rate $d\dot{N}/d\beta$ seen in the right panel of
Fig.~\ref{F:dNdbeta}.  The bends in these curves marking the threshold for full disruption in the empty loss-cone regime
migrate to lower values of $\beta$ consistent with the SBH mass dependence of the mapping $\beta_N(\beta)$ shown in
Fig.~\ref{F:betaN}.  The curves also end abruptly at $\beta_{\rm cap}$ as can be seen in the red curve corresponding to SBH
mass $M_\bullet = 10^7 M_\odot$ for which $\beta_{\rm cap} \simeq 1.27$.

\subsection{Distributions of peak fallback accretion rate $\dot{M}_{\rm peak}$ and time delay $t_{\rm peak}$}
\label{SS:dpeak}

\begin{figure*}[t]
\subfloat{\includegraphics[width=3.5in,height=2.5in]{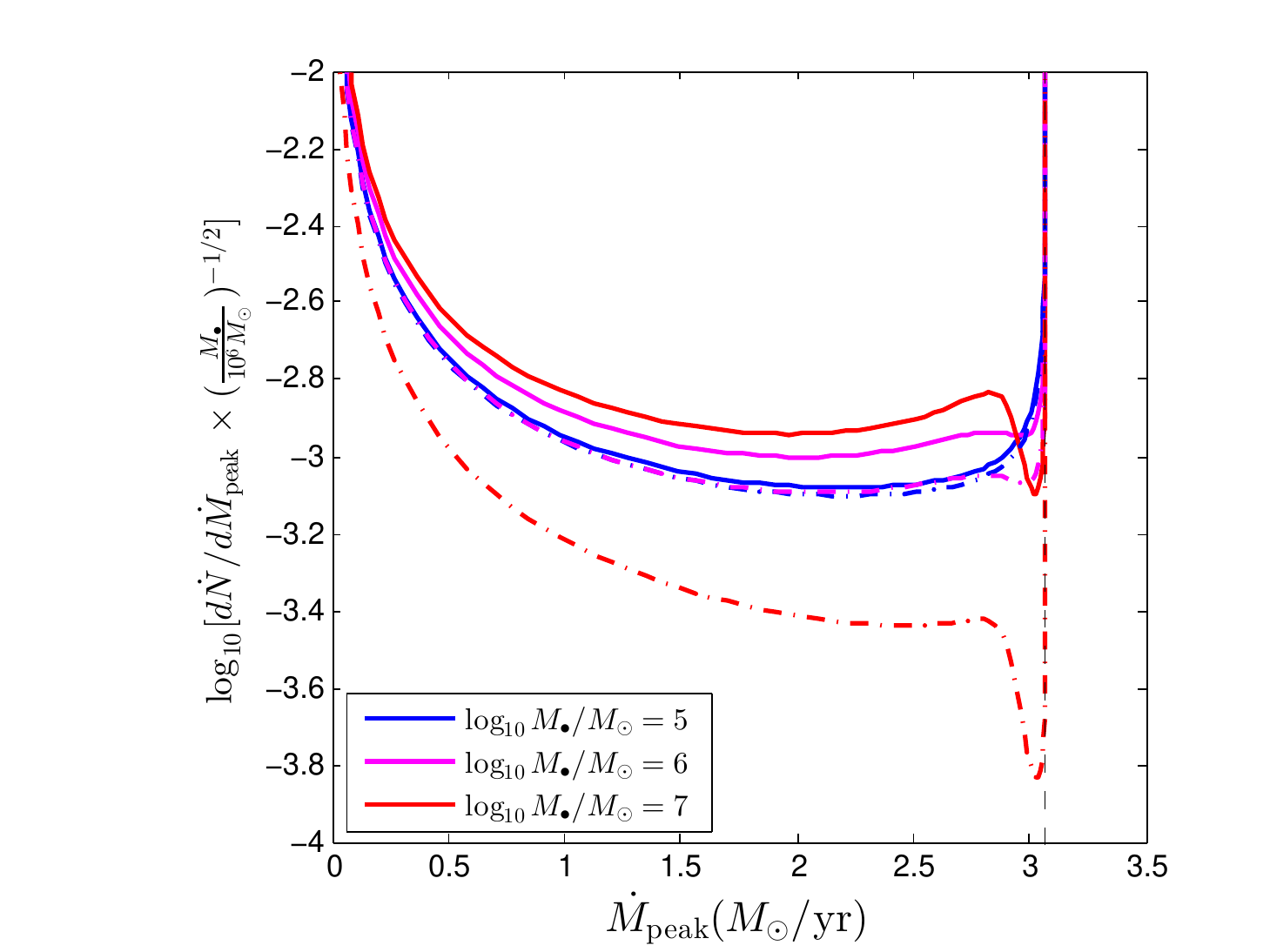}} 
\subfloat{\includegraphics[width=3.5in,height=2.5in]{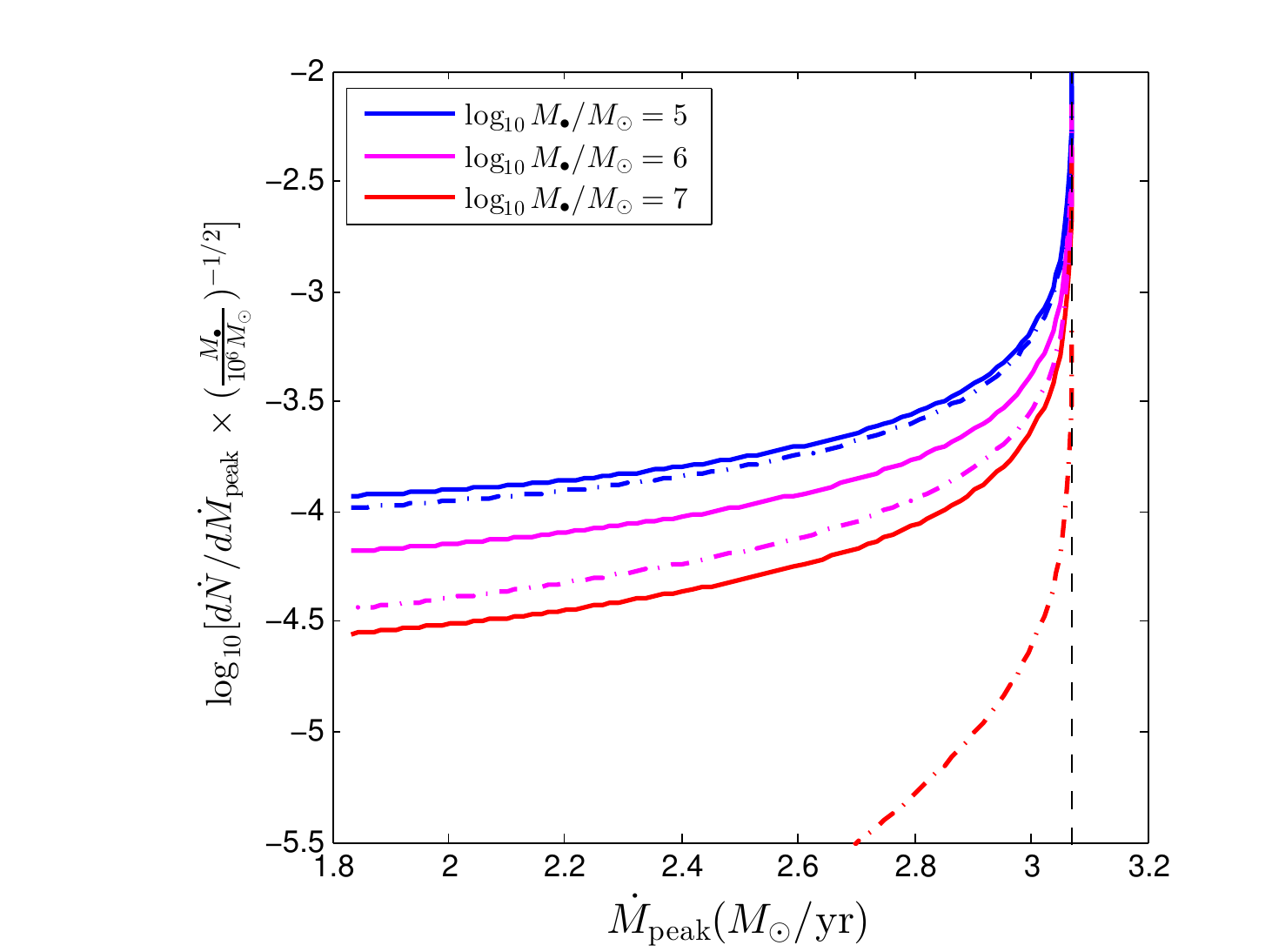}}\
\subfloat{\includegraphics[width=3.5in,height=2.5in]{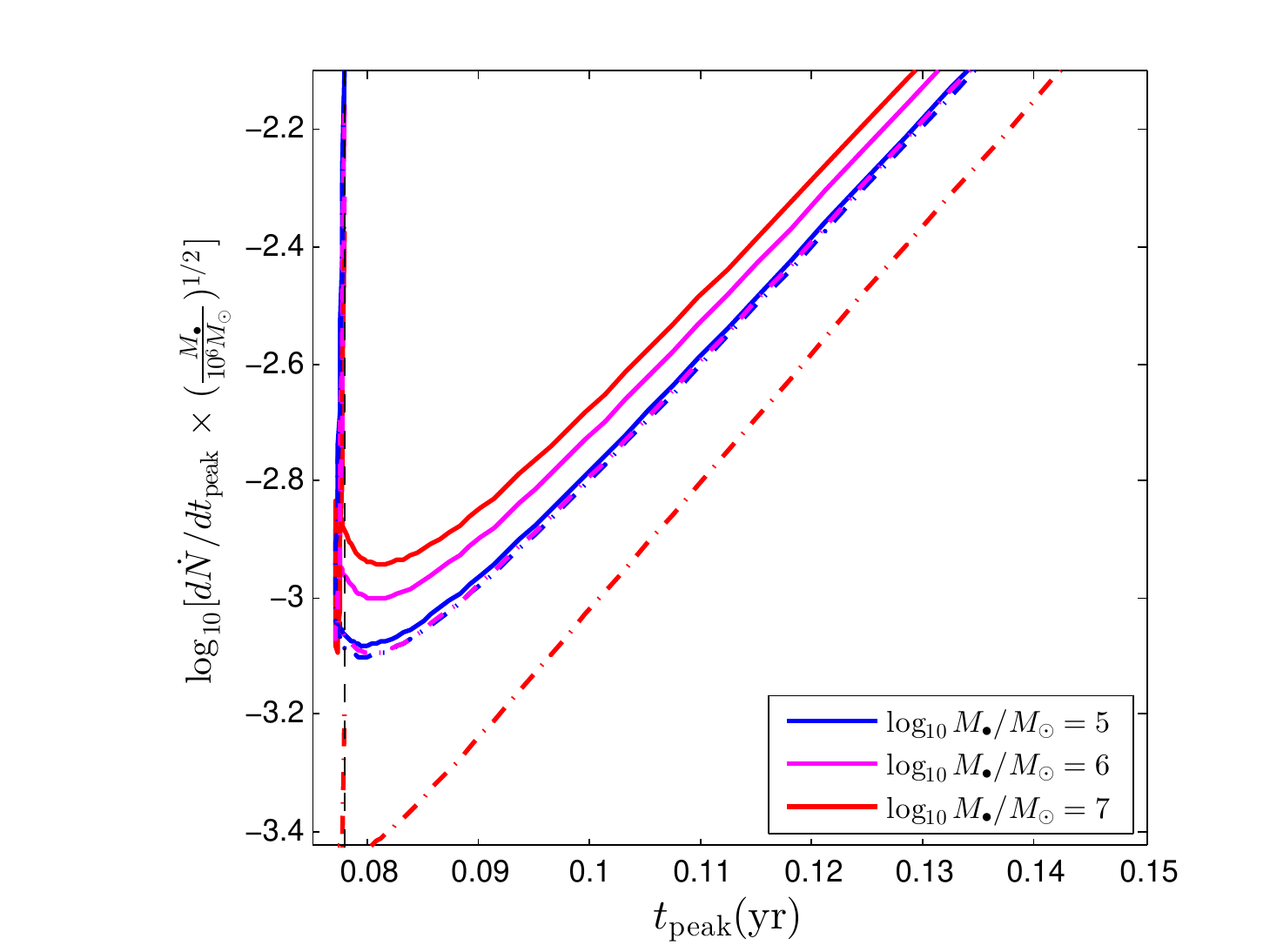}}
\subfloat{\includegraphics[width=3.5in,height=2.5in]{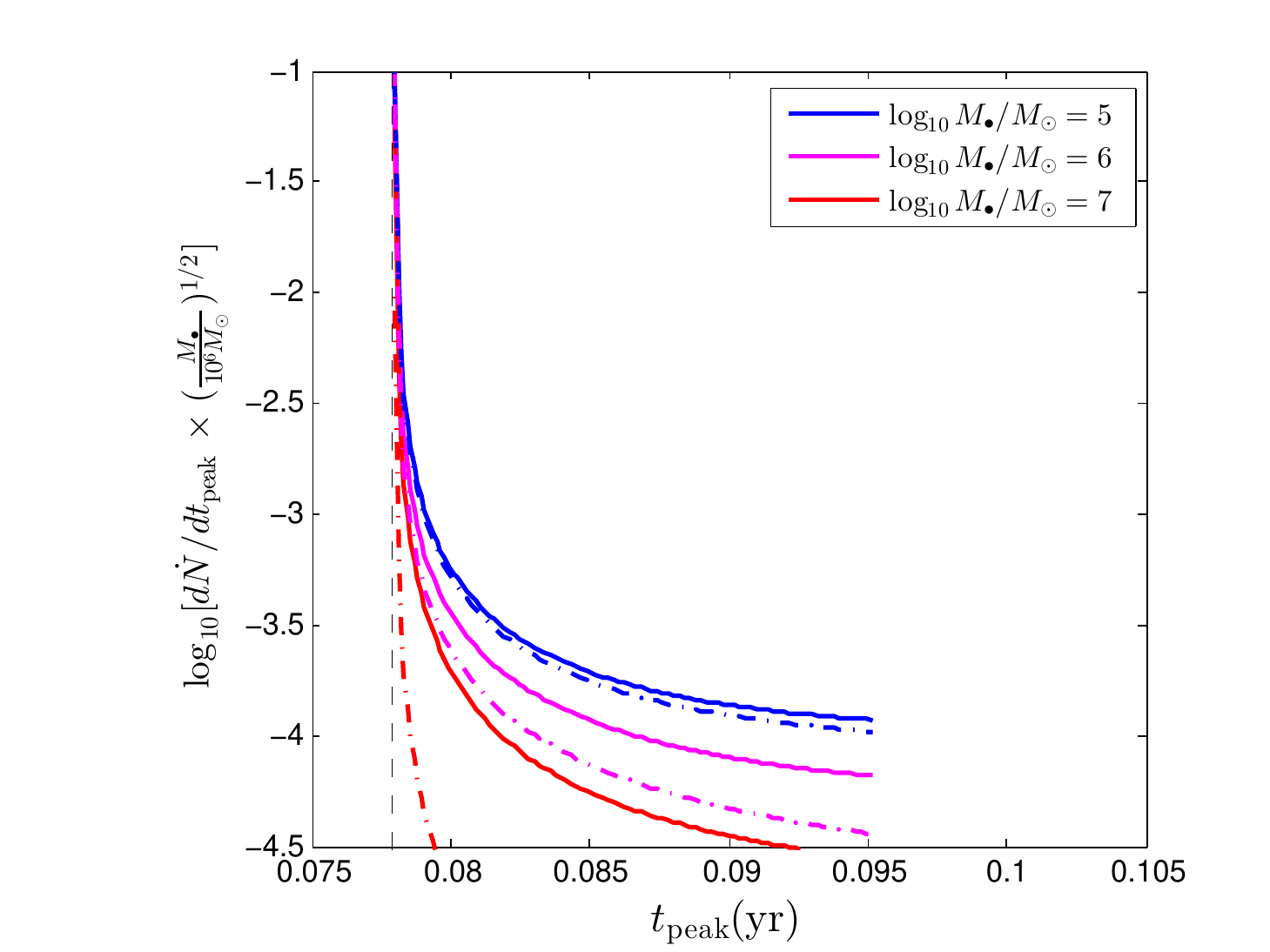}}
\caption{The differential TDE rates $d\dot{N}/d\dot{M}_{\rm peak}$ (top panels) and $d\dot{N}/dt_{\rm peak}$
(bottom panels) as functions of the peak fallback accretion rate $\dot M_{\rm peak}$ and time delay $t_{\rm peak}$ for partial
(left panels) and full (right panels) disruptions.  To clarify the presentation of relativistic effects, we have scaled out the
explicit dependence on SBH mass $M_\bullet$ given on the right-hand side of Eq.~(\ref{E:HydroFits}).  The solid
(dot-dashed) curves show the differential TDE rates in Newtonian gravity (general relativity). The vertical dashed black lines
indicate the values of $\dot M_{\rm peak}$ and time delay $t_{\rm peak}$ and the threshold $\beta_{N,d} = 1.9$ of full
disrpution.} \label{F:4rat}
\end{figure*}

Guillochon and Ramirez-Ruiz \cite{2013ApJ...767...25G} performed a series of Newtonian hydrodynamical simulations at
different penetration factors $\beta_N$ and used them to derive analytic fits to the peak fallback accretion rate
$d\dot{M}_{\rm peak}/dt$ and time delay $t_{\rm peak}$ for Solar-type stars with polytropic index $\gamma = 4/3$:
\begin{subequations} \label{E:HydroFits}
\begin{align}
\dot M_{\rm peak} &= A_{4/3} \left( \frac{M_\bullet}{10^6 M_\odot} \right)^{-1/2}, \\
t_{\rm peak} &= B_{4/3} \left( \frac{M_\bullet}{10^6 M_\odot} \right)^{1/2},
\end{align}
\end{subequations}
where,
\begin{subequations} \label{E:fitbeta}
\begin{align}
A_{4/3} &= {\rm exp}\left( \frac{27.3-27.5\beta_N +3.87\beta_N^2}{1-3.26\beta_N-1.39\beta_N^2} \right), \\
B_{4/3} &= \frac{-0.387+0.573\sqrt{\beta_N}-0.312\beta_N}{1-1.27\sqrt{\beta_N}-0.9\beta_N},
\end{align}
\end{subequations}
and $0.6 \leq \beta_N \leq 4.0$.  Given these fits, the mapping $\beta_N(\beta)$ derived in Sec.~\ref{S:map}, and the
differential TDE rate $d\dot{N}/d\beta$ obtained in the previous subsection, we can calculate differential rates
\begin{equation} \label{E:diffrateX}
\frac{d\dot{N}}{dX} = \frac{d\dot{N}}{d\beta} \left( \frac{d\beta_N}{d\beta} \frac{dX}{d\beta_N} \right)^{-1}
\end{equation}
for $X \in \{\dot{M}_{\rm peak},  t_{\rm peak} \}$.  We show these differential TDE rates in Fig.~\ref{F:4rat} below.  A
complication arises from the fact that the dependence on penetration factor in Eq.~(\ref{E:fitbeta}) is not monotonic;
$\dot M_{\rm peak}$ ($t_{\rm peak}$) increases~(decreases) with $\beta_N$ until the threshold for full disruption $\beta_{N,d}$ is
reached, then decreases (increases) for TDEs that penetrate more deeply into the SBH's potential well.  This $\beta_N$
dependence contradicts the naive predictions of freezing models which assume that the energy distribution of the tidal
debris is frozen in at the disruption radius $r_d$, but does not fatally compromise our analyis.  We address this complication
by plotting the TDEs from partial ($\beta_N < \beta_{N,d}$) and full ($\beta_N > \beta_{N,d}$) disruptions separately in
the left and right panels of Fig.~\ref{F:4rat}.  The differential rates diverge at the extremum $\beta_N = \beta_{N,d}$ shown by
the vertical black dashed lines in these plots, but the total rates (areas under the curves) remain finite.  We see that the
stronger tides of relativity suppress the rates of both full and partial TDEs, particularly for SBH masses
$M_\bullet \geq 10^7 M_\odot$ for which portions of phase space with $\beta_N > \beta_{N,d}$ lie primarily in the empty
loss-cone regime.  This emptiness even reduces the rate of partial TDEs because of the sharper gradients driving stronger
diffusion across the boundary as seen  by the dip near the threshold in the top left panel of Fig.~\ref{F:4rat}.

\subsection{Relativistic correction to the debris energy distribution} \label{SS:relE}

\begin{figure*}[t]
\subfloat{\includegraphics[width=3.5in,height=3in]{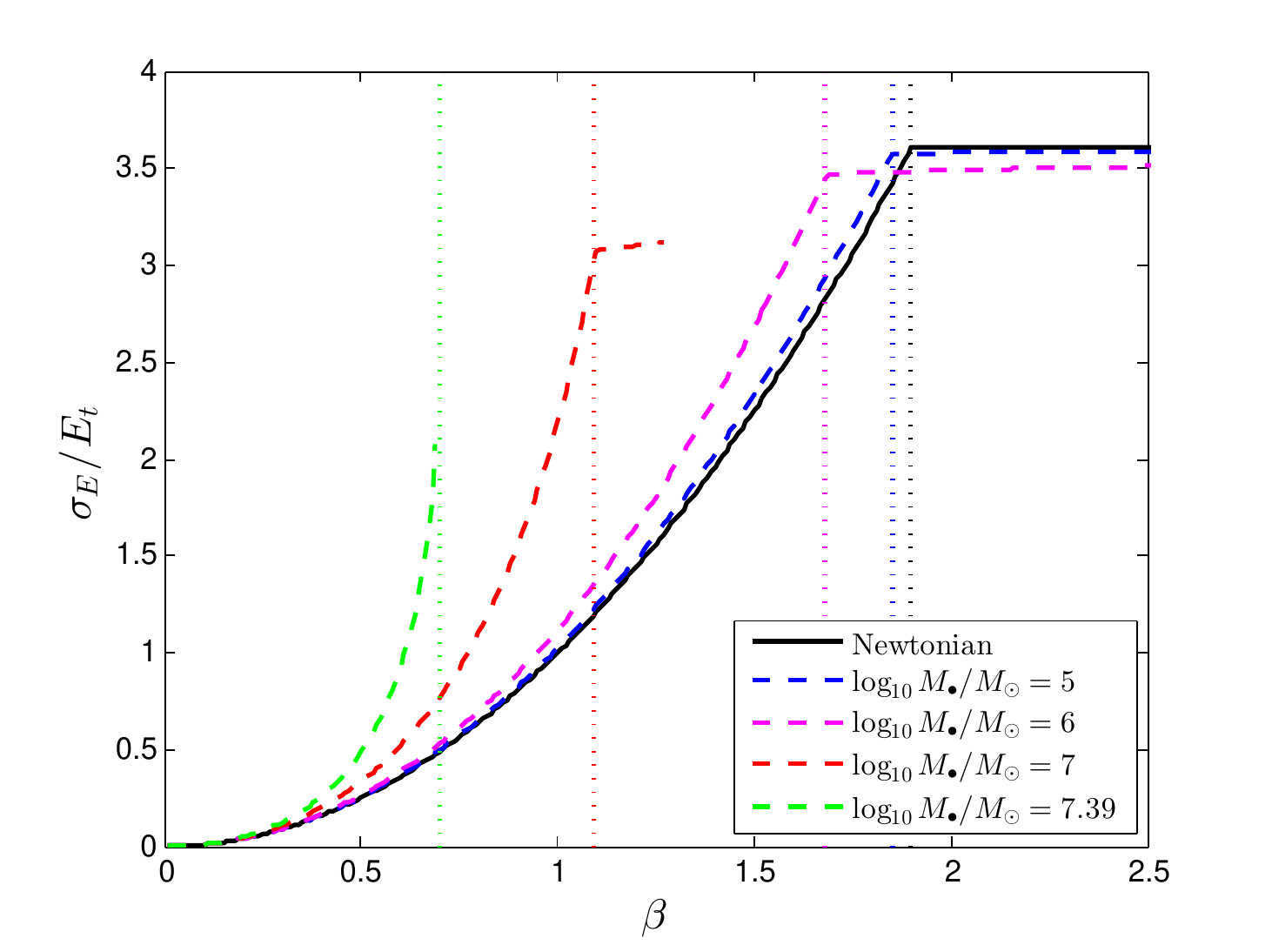}}
\subfloat{\includegraphics[width=3.5in,height=3in]{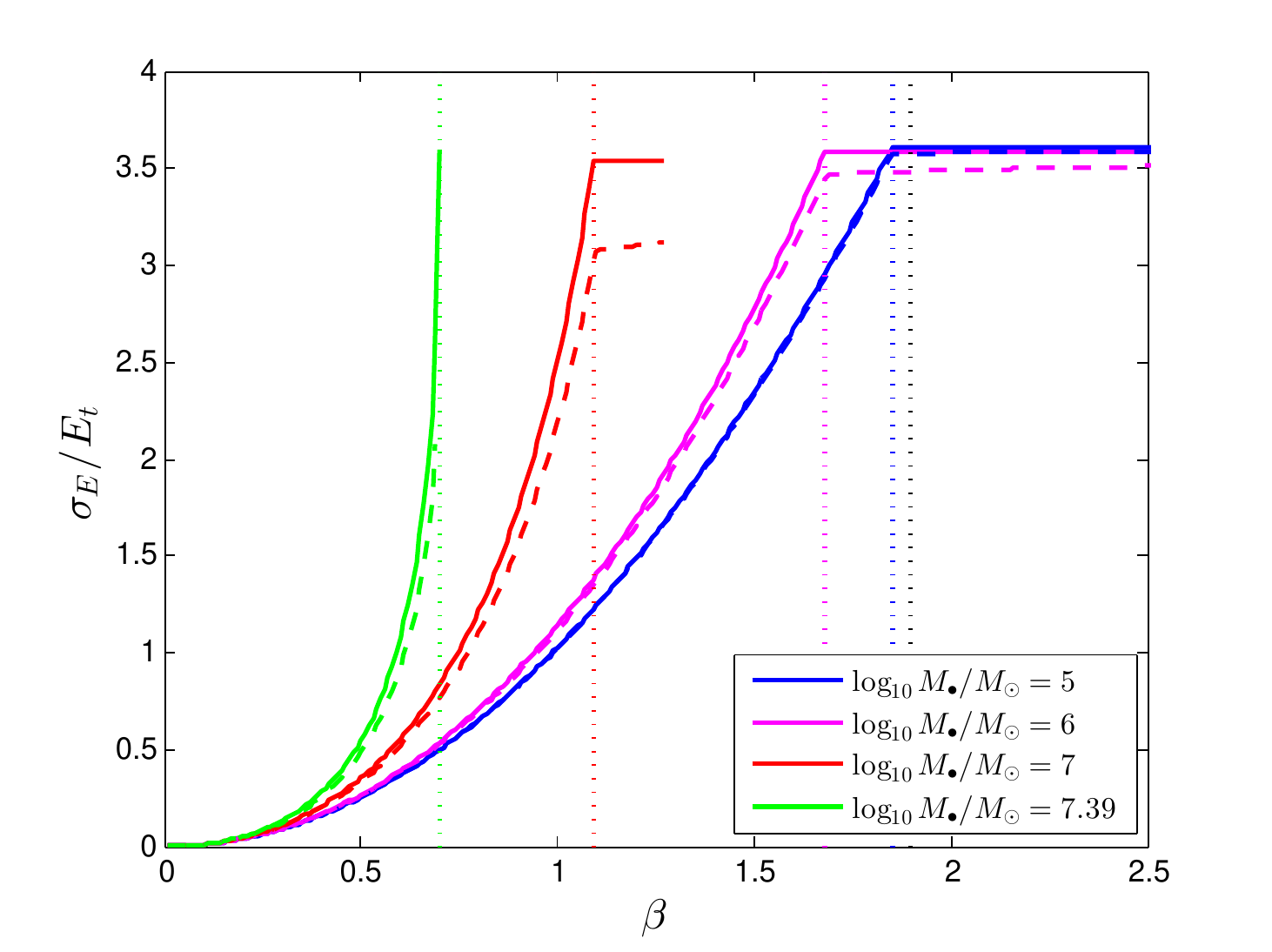}}
\caption{Dimensionless width $\sigma_E/E_t$ of the tidal debris energy distribution as a function of penetration factor
$\beta$.  The solid curves show the Newtonian predictions $\sigma_{E,\rm N}/E_t$, while the dashed blue, magenta, red,
and green curves show the relativistic predictions $\sigma_{E,\rm GR}/E_t$ for $M_\bullet/M_\odot = 10^5$, $10^6$,
$10^7$, and $10^{7.39}$.  The last value is $M_{\rm max}$, the most massive SBH capable of fully disrupting a Solar-type star. 
The vertical dotted lines show the thresholds $\beta_d$ for full disruption; each of the relativistic curves end at
$\beta_{\rm cap}$.  The single black curve in the left panel shows the Newtonian prediction in mapping (2) which is
independent of SBH mass, while the three solid curves in the right panel show the Newtonian predictions for mapping (3).}
\label{F:sigmaE23}
\end{figure*}

The estimated differential TDE rates $d\dot{N}/d\dot{M}_{\rm peak}$ and $d\dot{N}/dt_{\rm peak}$ in the previous subsection
accounted for relativistic corrections to the boundaries of the loss cone, but in the interest of a systematic exploration of
relativistic effects we did not simulataneously include relativistic corrections to the energy distribution of the tidal debris.  We
turn our attention to this effect in this subsection.

A key approximation often used in the analysis of TDEs is that the tidal debris travels on ballistic trajectories following
disruption, and that orbit circularization followed by viscous accretion occurs promptly after the bound debris falls back to
pericenter \cite{1988Natur.333..523R,1989Evans,1994ApJ...422..508K}.  The fallback accretion rate onto the SBH, and
thus its peak value $\dot{M}_{\rm peak}$ and the time $t_{\rm peak}$ between disruption and when this peak occurs, is
determined by the distribution of orbital periods of the tidal debris, which is in turn set by the energy distribution through
Kepler's third law
\begin{equation} \label{E:Kep3}
\tau = 2\pi \left( \frac{a^3}{M_\bullet} \right)^{1/2} = 2\pi M_\bullet (2E)^{-3/2}\, .
\end{equation}
This Newtonian relation remains an excellent approximation even for relativistic TDEs, because most of the debris is on
highly eccentric orbits prior to circularization and spends most of its time near apocenter where Eq.~(\ref{E:Kep3}) holds.
However, if the tidal disruption itself occurs near pericenter where relativistic effects are strongest, Newtonian predictions
for the specific binding energy $E$ entering into Eq.~(\ref{E:Kep3}) may not be accurate.  TDE simulations in general
relativity \cite{2013MNRAS.434..909H,2014PhRvD..90f4020C,2016MNRAS.455.2253B,2015ApJ...804...85S,2013PhRvD..87j4010C} naturally yield proper relativistic energy distributions, but our goal in this section is to study how relativistic
corrections might alter the predictions of Newtonian simulations.

We addressed this issue in Kesden \cite{Mike2nd}, where we found that for an undistorted star of radius $R_\star$, the width
of the potential-energy distribution before disruption and thus the width of the debris energy distribution after disruption is
given in general relativity by the expression
\begin{equation} \label{E:relE}
\sigma_{E,\rm GR} = |\lambda^\alpha_{~(i)} \nabla_\alpha E| R_{\star} = |g_{\beta \gamma}
\lambda^\beta_{~(0)} \lambda^\alpha_{~(i)} \Gamma^\gamma_{\alpha t}| R_\star,
\end{equation}
where $g_{\beta\gamma}$ is the metric tensor, $\Gamma^{\gamma}_{\alpha t}$ are the Christoffel symbols, and
$\lambda^\beta_{~(0)}$ and $\lambda^\alpha_{~(i)}$ are the orthonormal tetrad of basis 4-vectors determined by our choice
of Fermi normal coordinates in Sec~\ref{S:rel}.  For $\beta < \beta_d$ corresponding to partial disruptions, tidal debris is
assumed to be liberated at pericenter where the tidal forces are strongest.  Evaluating Eq.~(\ref{E:relE}) at pericenter for
such orbits yields
\begin{align}
\sigma_{E,\rm GR, PD} &= \frac{M_{\bullet}R_{\star}}{r^2} \left( 1 - \frac{2M_\bullet}{r}  \right)^{-1/2} \notag \\
&=  E_t \left( \frac{r_t}{r} \right)^2 \left( 1 - \frac{2M_\bullet}{r}  \right)^{-1/2}. \label{E:relEPD}
\end{align}
The pericenter $r$ in Boyer-Lindquist coordinates can be expressed in terms of the angular momentum $L$ using
Eq.~(\ref{E:rpL}),
\begin{equation} \label{E:Etid}
E_t \equiv \frac{M_\bullet R_\star}{r_t^2} = \left( \frac{M_\bullet}{M_\star} \right)^{1/3} \frac{M_\star}{R_\star} =
\left( \frac{M_\bullet}{M_\star} \right)^{1/3} E_\star
\end{equation}
is an order-of-magnitude estimate for the width of this energy distribution, and $E_\star \equiv M_\star/R_\star$ is a similar
estimate for the specific self-binding energy of the star.  The mass hierarchy $M_\star \ll M_\bullet$ implies that
$E_\star \ll E_t$, supporting the assumption of freezing models that the relative velocities of fluid elements at the time of
disruption can be neglected when determining the debris energy distribution.

In trying to compare this relativistic result to Newtonian predictions, we encounter the same issue of choosing a mapping
between the two gravitational theories that we addressed in Sec.~\ref{S:map}.  Mapping (1), which was used in Kesden
\cite{Mike2nd}, compared Eq.~(\ref{E:relEPD}) to Newtonian orbits with the same pericenter coordinate and found
\begin{equation} \label{E:NewtEPD1}
\sigma_{E,\rm N, PD(1)} = \frac{M_\bullet R_\star}{r^2} = E_t \left( \frac{r_t}{r} \right)^2.
\end{equation}
Dividing Eq.~(\ref{E:relEPD}) by Eq.~(\ref{E:NewtEPD1}) yields a peak correction of $\sqrt{2}$ for $r = 4M_\bullet$, the
minimum pericenter for an orbit that avoids direct capture by the horizon.  As argued in Sec.~\ref{S:map} however, this
mapping is a somewhat unnatural way to identify orbits in the two theories. 

Mapping (2) compared Schwarzschild geodesics to Newtonian orbits with the same angular momentum $L$ (and hence
penetration factor $\beta = L_t^2/L^2$).  For this choice, the width of the Newtonian energy distribution is
\begin{equation} \label{E:NewtEPD2}
\sigma_{E,\rm N, PD(2)} = \frac{M_\bullet R_\star}{r_L^2} = E_t \left( \frac{2M_\bullet}{L^2} \right)^2 = \beta^2 E_t~,
\end{equation}
where $r_L$ is the pericenter of a Newtonian orbit with angular momentum $L$.  Freezing models posit that the energy
distribution is frozen in for $\beta > \beta_{N,d}$, implying that $\sigma_{E,\rm N, FD(2)} = \beta_{N,d}^2 E_t$ for full
disruptions in mapping (2).  Although there is no nice analytic expression for the widths $\sigma_{E,\rm GR, FD}$ for full
disruption in general relativity, they can be calculated numerically using Eq.~(\ref{E:relE}).  We show the widths of these
distributions in mapping (2) for both partial and full disruptions in the left panel of Fig.~\ref{F:sigmaE23}.  We see that for
partial disruptions $\beta < \beta_d$, the energy distribution is broader in relativity than Newtonian gravity.  The ratio
$\sigma_{E,\rm GR}/\sigma_{E,\rm N(2)}$ reaches a maximum value of $4\sqrt{2}$ at $M_{\rm max}$, even larger than the
peak correction in mapping (1).  However, the weaker tides in Newtonian gravity imply that stars can reach larger
penetration factors $\beta_{N,d} > \beta_d$ before being fully disrupted.  SBHs with masses
$M_\bullet \simeq 10^6 M_\odot$ can have deeply penetrating encounters $\beta_d < \beta < \beta_{\rm cap}$ where the
star still manages to avoid direct capture.  For such penetration factors, the relativistic prediction can fall below that in
Newtonian gravity as can be seen near the right edge of the left panel of Fig.~\ref{F:sigmaE23}.  It is interesting to note that
the relativistic predictions retain some mild $\beta$ dependence for full disruptions $\beta > \beta_d$, unlike in Newtonian
freezing models.  This is because the tidal tensor $C^{(i)}_{\quad(j)}$ and energy width $\sigma_{E,\rm GR}$ are
velocity-dependent in relativity, even if evaluated at the fixed tidal acceleration $2M_\star/(\eta R_\star)^2$ set by the star's
self gravity.

\begin{figure}[t!]
\centering
\includegraphics[width=3.5in,height=3.25in]{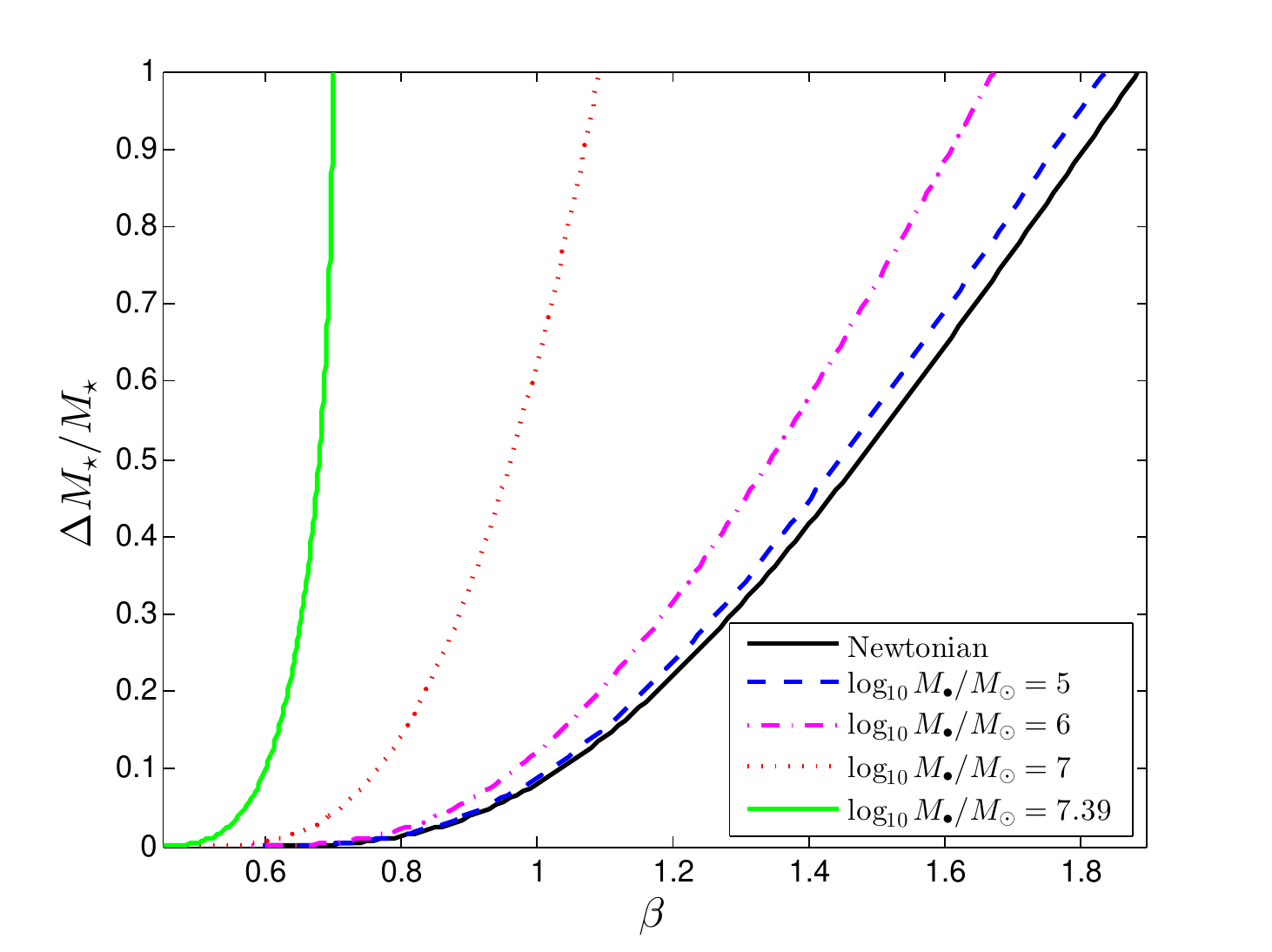}
\caption{The fraction $\Delta M_\star/M_\star$ of the stellar mass lost in partial disruptions as a function of penetration
factor $\beta$.  The solid black curve shows a fit to Newtonian simulations \cite{2013ApJ...767...25G}, while the blue,
magenta, red, and green curves show the corresponding relativistic predictions made using our mapping $\beta_N(\beta)$
for SBH masses $M_\bullet/M_\odot = 10^5$, $10^6$, $10^7$, and $M_{\rm max}/M_\odot \simeq 10^{7.39}$.}
\label{F:frac}
\end{figure}

Mapping (3) solves the problem of stars being fully disrupted at different values of $\beta$ in the two theories by identifying
the Schwarzschild geodesic with penetration factor $\beta$ with the Newtonian orbit with penetration factor $\beta_N(\beta)$
on which a star experiences the same peak tidal force at pericenter as shown in Fig.~\ref{F:betaN}.  In this case, the width
of the energy distribution for partial disruptions is
\begin{equation} \label{E:NewtEPD3}
\sigma_{E,\rm N, PD(3)} = \frac{M_\bullet R_\star}{r_F^2} = \beta_N^2(\beta) E_t
= \frac{M_{\bullet}R_{\star}}{r^2} \left( \frac{r + M_\bullet}{r - 2M_\bullet}  \right)^{2/3}~,
\end{equation}
where $r_F$ is the pericenter of a Newtonian orbit with penetration factor $\beta_N(\beta)$ and $r$ as previously is the
pericenter in Boyer-Lindquist coordinates of a Schwarzschild geodesic with penetration factor $\beta$.  This width again
freezes out at $\sigma_{E,\rm N, FD(3)} = \beta_{N,d}^2 E_t$ for full disruptions as in mapping (2), but these full disruptions
now occur for $\beta > \beta_d$ as in general relativity.  We show $\sigma_{E,\rm GR}$ and $\sigma_{E,\rm N(3)}$ as
functions of penetration factor $\beta$ in the right panel of Fig.~\ref{F:sigmaE23}.  We see that full disruption occurs at
$\beta_d$ in both theories with this mapping, but that for a given penetration factor $\beta$, the relativistic predictions
$\sigma_{E,\rm GR}$ are now below the Newtonian predictions $\sigma_{E,\rm N(3)}$.  The deeper penetration needed for
tidal disruption in Newtonian gravity leads to steeper potential gradients and thus broader tidal debris energy distributions.
The ratio $\sigma_{E,\rm GR}/\sigma_{E,\rm N(3)}$ reaches a minimum value of $(128/625)^{1/6} \simeq 0.768$ at
$M_{\rm max}$.  The Newtonian orbit mapped to the Schwarzschild geodesic with $L_{\rm cap} = 4M_\bullet$ has
$L_N =(1024/5)^{1/6}M_\bullet \simeq 2.43 M_\bullet < L_{\rm cap}$, but this is not a problem in principle since Newtonian
point masses have no horizons to capture stars.

Which of the three expressions given by Eqs.~(\ref{E:NewtEPD1}), (\ref{E:NewtEPD2}), (\ref{E:NewtEPD3}) is the "right" one
to use when comparing TDEs in Newtonian gravity with those in general relativity?  The answer to this question depends on
which of the three mappings discussed in Sec.~\ref{S:map} you are using to relate orbits in the two theories.  If you want to
compare orbits with the same angular momentum $L$, Eq.~(\ref{E:NewtEPD2}) provides the width of the tidal energy
distribution for a Newtonian orbit with the same penetration factor $\beta$, leading to more tightly bound debris in relativity
for partial disruptions as seen in the left panel of Fig.~\ref{F:sigmaE23}.  However, this choice implies that different fractions
of the stellar mass are stripped away by tides as seen in Fig.~\ref{F:frac}.  More material is lost by the partially disrupted star
in relativity, further enhancing the fallback accretion rate in addition to the more tightly bound material falling back more
quickly.  For $\beta_d < \beta < \beta_{d,N}$, full disruptions occur in general relativity but not Newtonian gravity, while for
$\beta > \beta_{d,N}$ stars are fully disrupted in both theories but the tidal debris is more tightly bound in Newtonian gravity.

\begin{figure*}[t]
\subfloat{\includegraphics[width=3.5in,height=3in]{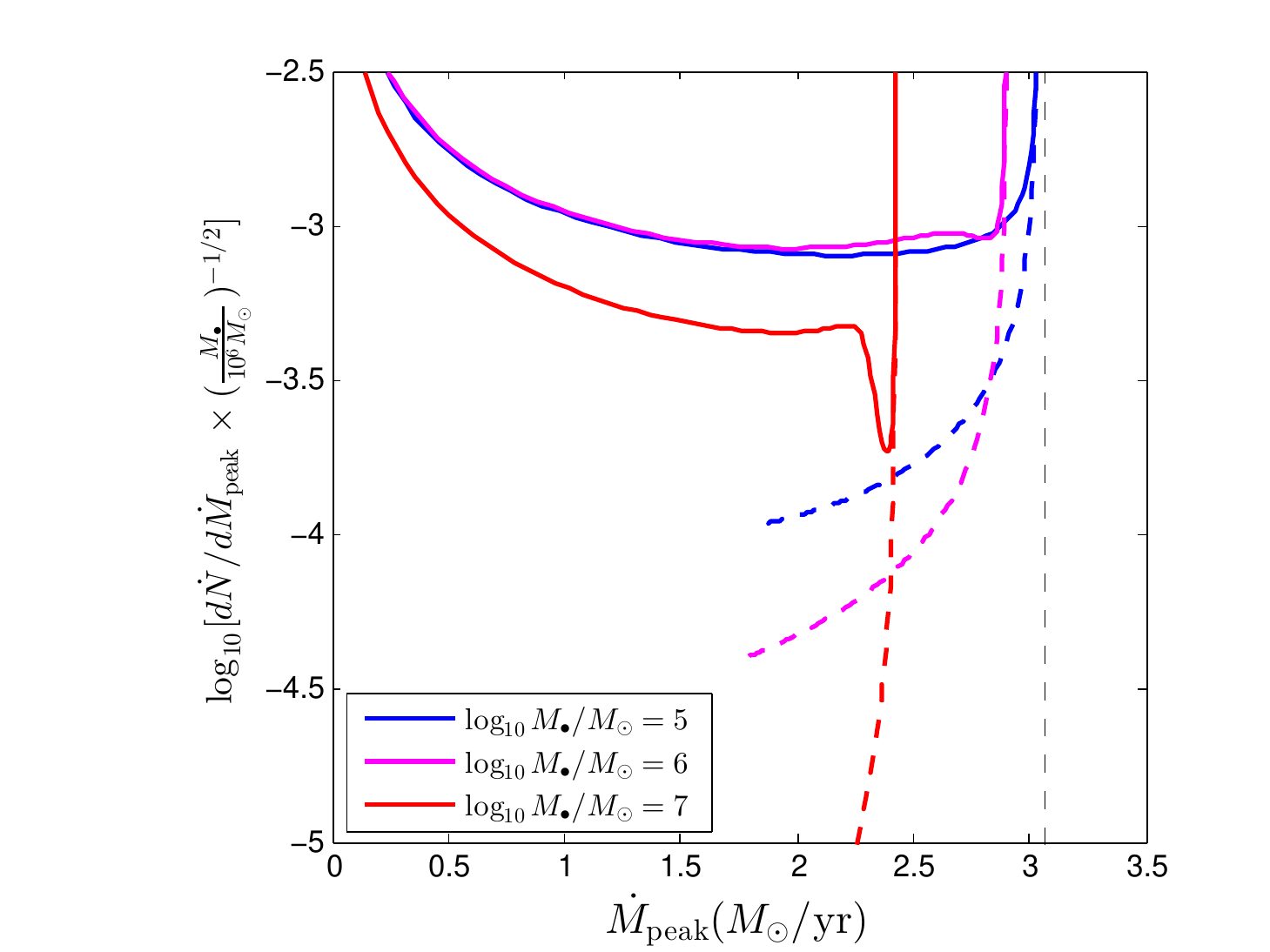}}
\subfloat{\includegraphics[width=3.5in,height=3in]{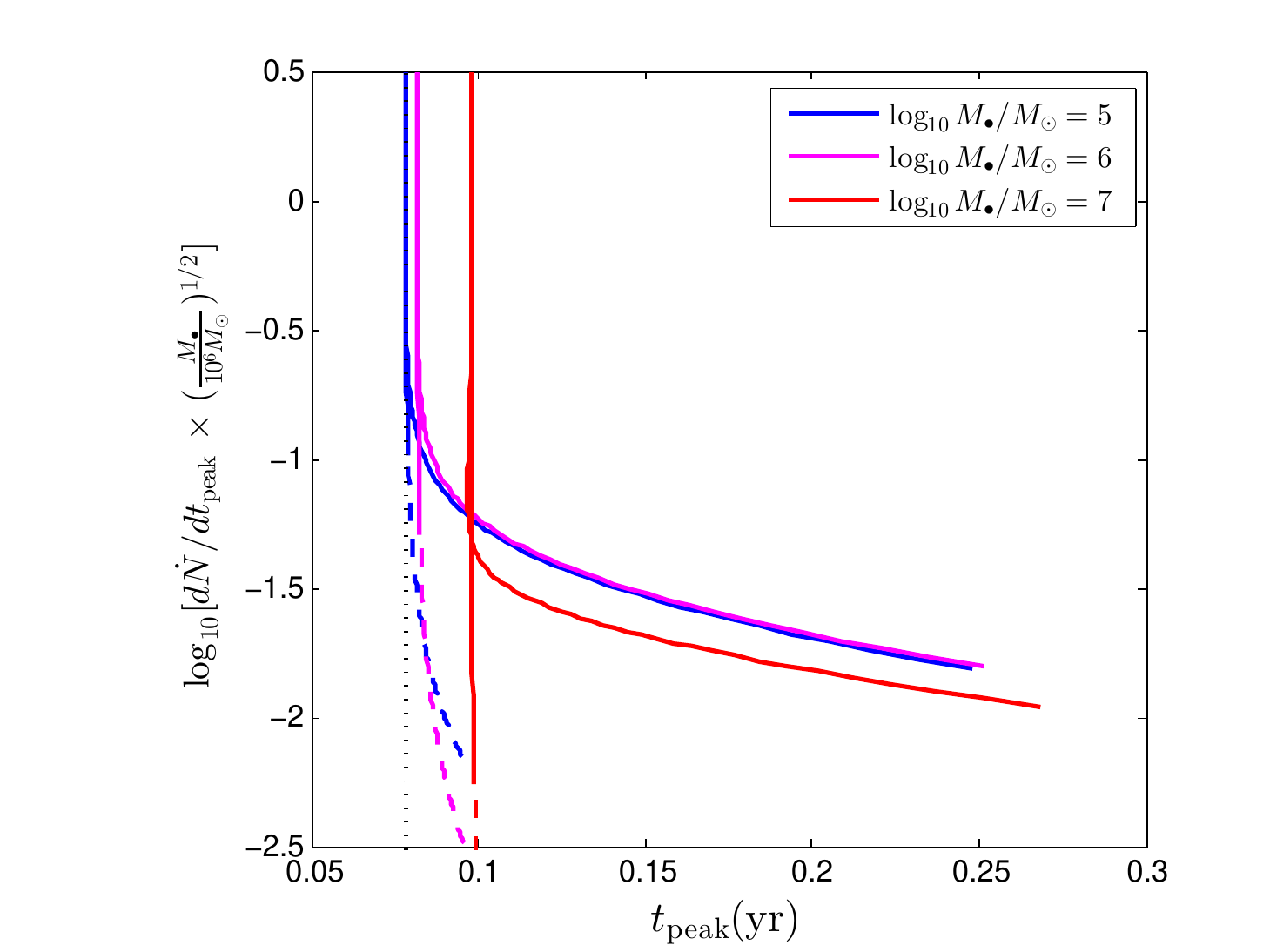}}
\caption{The differential TDE rates $d\dot N/d\dot{M}_{\rm peak}$ (left panel) and $d\dot N/dt_{\rm peak}$ (right panel) as
functions of the peak fallback accretion rate $\dot{M}_{\rm peak}$ and time delay $t_{\rm peak}$ between disruption and
when this peak is reached.  The blue, magenta, and red curves show SBH masses $M_\bullet/M_\odot = 10^5$, $10^6$,
and $10^7$.  The solid curves show partial disruptions while the dashed curves show full disruptions.  The vertical black
dotted lines show the values in Newtonian gravity at the threshold $\beta_{N,d}$ for full disruption.}  \label{F:2rat}
\end{figure*}

If you are instead trying use Newtonian TDE simulations like those in Guillochon and Ramirez-Ruiz
\cite{2013ApJ...767...25G} to predict what this process might be like in relativity, you should use mapping (3) to relate orbits
experiencing the same peak tidal forces and thus the same amount of mass loss by partially disrupted stars.
Eq.~(\ref{E:NewtEPD3}) should be used to determine the relativistic correction $\sigma_{E,\rm GR}/\sigma_{E,\rm N(3)}$ as
shown in the right panel of Fig.~\ref{F:sigmaE23}.  We use this correction to predict the peak fallback accretion rate and time
delay
\begin{subequations} \label{E:RelObs}
\begin{align}
\dot M_{\rm peak}(\beta) &= \left( \frac{\sigma_{E,\rm GR}}{\sigma_{E,\rm N(3)}} \right)^{3/2}
\times \dot{M}_{\rm peak}[\beta_N(\beta)]\,, \\
t_{\rm peak}(\beta) &= \left( \frac{\sigma_{E,\rm GR}}{\sigma_{E,\rm N(3)}} \right)^{-3/2}
\times t_{\rm peak}[\beta_N(\beta)]\,,
\end{align}
\end{subequations}
in general relativity, where the exponent of 3/2 follows from the energy dependence in Kepler's third law (\ref{E:Kep3}).  We
show the corrected differential TDE rates $d\dot N/d\dot{M}_{\rm peak}$ and $d\dot N/dt_{\rm peak}$ in
Fig.~\ref{F:2rat}.  We see that the peak fallback accretion rates $\dot{M}_{\rm peak}$ have been reduced below their
Newtonian values by $\sim 20\%$ for a $10^7 M_\odot$ SBH, consistent with Eq.~(\ref{E:RelObs}) and the ratios of the
curves shown in the right panel of Fig.~\ref{F:sigmaE23}.  The time delays $t_{\rm peak}$ show a corresponding increase.
Fig.~\ref{F:2rat} conveys the qualitative message of this subsection: the stronger tides in general relativity allow SBHs to
tidally disrupt stars at lower penetration factors $\beta$ leading to less tightly bound debris and lower fallback accretion
rates.

\subsection{Capture of tidal debris by the event horizon} \label{SS:relL}

In this subsection, we focus on an issue that was only briefly addressed in Kesden \cite{Mike2nd}: tidal debris can be
captured by the event horizon of an SBH even if the disrupted star is not initially on a capture orbit.  In general relativity, the
tidal debris will have a distribution of specific angular momentum $L$ with a width given by 
\begin{equation} \label{E:relL}
\sigma_{L, {\rm GR}} = |\lambda^\alpha_{~(i)}\nabla_{\alpha}L| R_\star =
|g_{\beta \gamma} \lambda^\beta_{~(0)} \lambda^\alpha_{~(i)} \Gamma^{\gamma}_{\alpha \phi}| R_\star
\end{equation}
analogous to Eq.~(\ref{E:relE}) giving the width of the specific energy distribution.  This width is set at pericenter $r$ for
partial disruptions, allowing us to evaluate Eq.~(\ref{E:relL}) as
\begin{equation} \label{E:relLPD}
\sigma_{L, {\rm GR, PD}} = R_\star \left( \frac{2M_\bullet}{r} \right)^{1/2}
= L_t \left( \frac{M_\star}{M_\bullet} \right)^{1/3} \left( \frac{r_t}{r} \right)^{1/2}.
\end{equation}
The mass hierarchy $M_\star \ll M_\bullet$ between the star and SBH implies that $\sigma_{L, {\rm GR}} \ll L_t$ and thus that
it is usually a good approximation to assume that the specific angular momentum of the tidal debris is equal to that of the
initial star.  This contrasts with the debris energy distribution for which $E_t \gg \sigma^2$ implying that the tidal debris is
much more tightly bound to the SBH than the initial star.  However, for highly relativistic TDEs, the small amount of specific
angular momentum lost in the disruption process may be enough for some of the tidal debris to be captured by the horizon.
We examine this possibility by defining the dimensionless parameter
\begin{equation} \label{E:fL}
f_L \equiv \frac{L+\Delta L-L_{\rm cap}}{|\Delta L|} = \frac{L- 4M_\bullet}{|\Delta L|} - 1\, ,
\end{equation}
where $L$ is the specific angular momentum of the initial star, $\Delta L < 0$ is the change in specific angular momentum of
a fluid element in the disruption process, and $L_{\rm cap} = 4M_\bullet$ is the specific angular momentum threshold for
direct capture by the horizon.  An element of the tidal debris with $f_L < 0$ will be captured by the SBH, and $f_L = -1$ is
the minimum value of this parameter for a star that is not originally on a capture orbit ($L > L_{\rm cap}$).

\begin{figure*}[t]
\subfloat{\includegraphics[width=3.75in,height=3.3in]{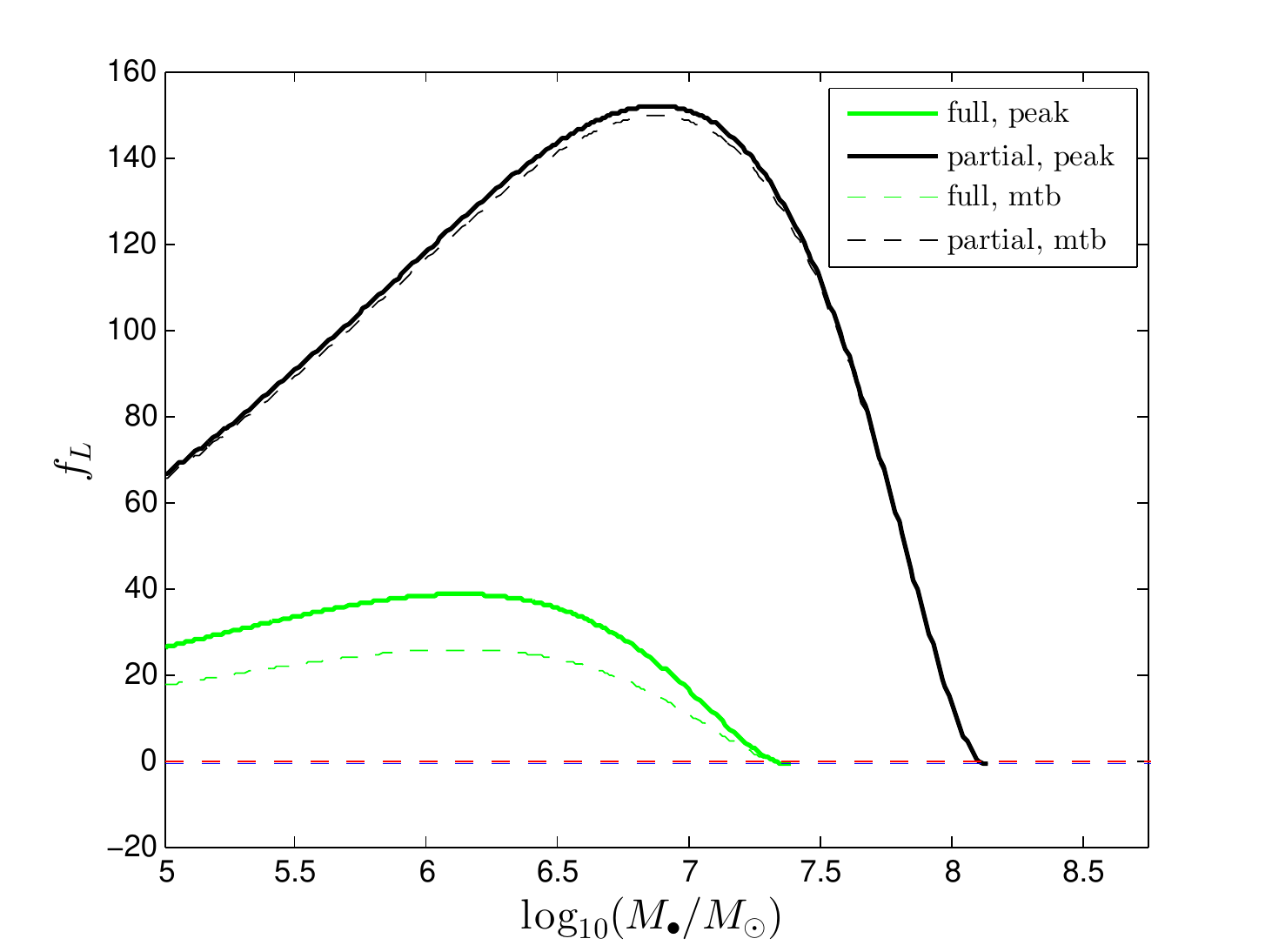}}
\subfloat{\includegraphics[width=3.75in,height=3.3in]{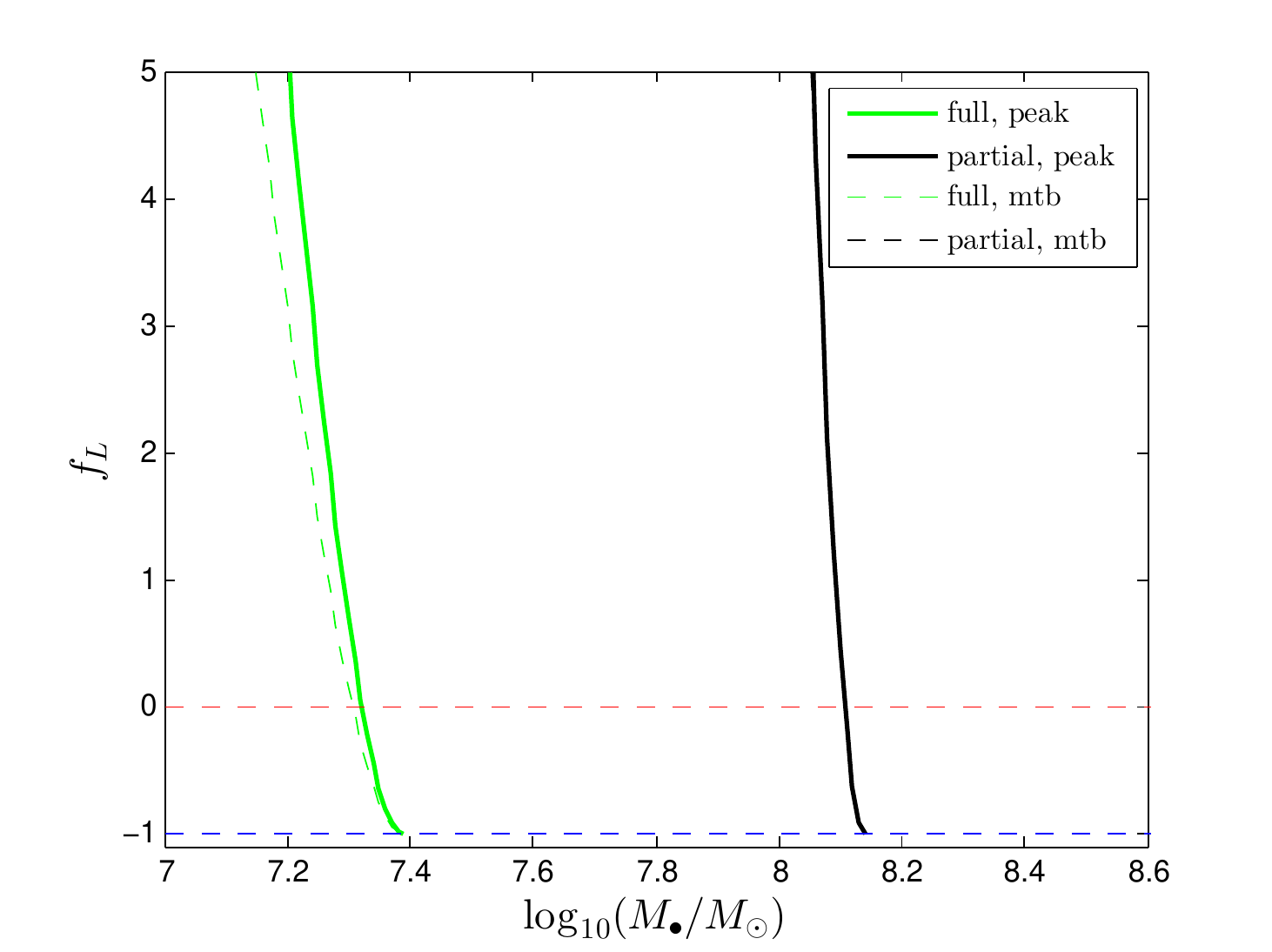}}
\caption{The parameter $f_L \equiv (L + \Delta L - L_{\rm cap})/|\Delta L|$, a measure of whether tidal debris is at risk of
direct capture by the event horizon, as a function of SBH mass $M_\bullet$.  The green (black) curves correspond to stars at
the thresholds $\beta = \beta_d~(\beta_{pd})$ for full (partial) disruptions.  The solid curves correspond to elements of the
tidal debris that fall back onto the SBH when fallback accretion peaks, while the dashed curves represent the most tightly
bound (mtb) elements (first to fall back).  The red (blue) horizontal dashed lines show the reference values $f_L = 0~(-1)$.}
\label{F:fL}
\end{figure*}

We need an estimate for $\Delta L$ to evaluate our new parameter $f_L$.  In the freezing model, an element of the tidal
debris located at $X^{(i)}$ in Fermi normal coordinates at tidal disruption has its specific energy and angular momentum
changed by an amount
\begin{subequations} \label{eq:deltaEL}
\begin{align}
\Delta E &\equiv X^{(i)} r_{E(i)} = X^{(i)} \lambda^{\alpha}_{~(i)}\nabla_{\alpha}E\, , \\
\Delta L &\equiv X^{(i)} r_{L(i)} = X^{(i)} \lambda^{\alpha}_{~(i)}\nabla_{\alpha}L\, ,
\end{align}
\end{subequations}
in the tidal-disruption process.  The element of the star that will become the most tightly bound element of the tidal debris
will have $X^{(i)}$ of magnitude $R_\star$ anti-aligned with $r_{E(i)}$, leading to the specific binding energy
$\sigma_{E, {\rm GR}}$ given by Eq.~(\ref{E:relE}).  The element of the star that loses the most angular momentum in the
tidal-disruption process will have $X^{(i)}$ of magnitude $R_\star$ anti-aligned with $r_{L(i)}$ and have its angular
momentum reduced by an amount $\sigma_{L, {\rm GR}}$ given by Eq.~(\ref{E:relL}).  If $r_{E(i)}$ and $r_{L(i)}$ are
parallel to each other (as at pericenter in Newtonian gravity, where both point in the radial direction), the same element of
the tidal debris will be both the mostly tightly bound (the first to fall back onto the SBH) and have lost the most angular
momentum (and thus be at greatest risk for direct capture).  We will assume this to be true, as it is exactly true in general
relativity at pericenter and a good approximation for those fluid elements with $f_L \lesssim 1$ most at risk of direct capture.
Under this assumption, TDEs for which $f_L \leq 0$ for $\Delta L = -\sigma_{L, {\rm GR}}$ in Eq.~(\ref{E:fL}) will have
suppressed fallback accretion rates and greater delays between disruption and the beginning of emission because direct
capture will have removed the most tightly bound debris.  Furthermore, if $r_{E(i)}$ and $r_{L(i)}$ are parallel, there is a 
one-to-one relationship
\begin{equation} \label{E:dLdE}
\Delta L = \frac{|r_{L(i)}|}{|r_{E(i)}|} E
\end{equation}
between the specific binding energy $E$ of an element of tidal debris and the specific angular momentum $\Delta L$ it loses
during tidal disruption.  If we use Kepler's third law (\ref{E:Kep3}) to relate the time delay $t_{\rm peak}$ to the specific
binding energy $E_{\rm peak}$ of debris accreted at that time, we can use Eq.~(\ref{E:dLdE}) to estimate $\Delta L$ and thus
$f_L$ for such debris.  When this estimate of $f_L$ becomes negative, all of the debris accreted before the fallback accretion
rate reaches its peak will be captured by the horizon.

We show $f_L$ for both peak and most tightly bound fluid elements as a function of SBH mass $M_\bullet$ in Fig.~\ref{F:fL},
for stars with penetration factors $\beta_d$ and $\beta_{pd}$ corresponding to the thresholds for full disruption and nonzero
partial disruption.  For the most tightly bound elements with $\Delta L = -\sigma_{L, {\rm GR}}$, Eq.~(\ref{E:fL}) yields
\begin{equation} \label{E:fLmtb}
f_L = \left( \frac{M_\bullet}{M_\star} \right)^{1/3} \left( \frac{r}{r_t} \right)^{1/2}
\left[ \frac{1}{\sqrt{\beta}} - \left( \frac{M_\bullet}{M_\star} \right)^{1/3} \sqrt{8E_\star} \right] - 1\,.
\end{equation}
For $M_\star \ll M_\bullet$, the term in the square brackets goes to $\beta^{-1/2}$ and $f_L \propto M_\bullet^{1/3}$, however
as $M_\bullet$ increases this term decreases and ultimately reaches zero ($f_L =-1$) for $L = L_{\rm cap}$ when the entire
star is directly captured.  This increase then subsequent decrease in $f_L$ as a function of $M_\bullet$ is seen in the left
panel of Fig.~\ref{F:fL} indicating that tidal debris is safe from direct capture for all but the most massive SBHs.  Only when
$M_\bullet$ is within $\sim 20\%$ of $M_{\rm max}$ does the direct capture of tidal debris become significant as seen in the
right panel of Fig.~\ref{F:fL}.  Although such events constitute only a small fraction of the total TDE rate, they are the TDEs
subject to the most extreme relativistic effects such as the pericenter precession that will be considered in the next
subsection.

\subsection{Relativistic pericenter precession} \label{SS:omega}

\begin{figure}[t]
\centering
\includegraphics[width=0.52\textwidth]{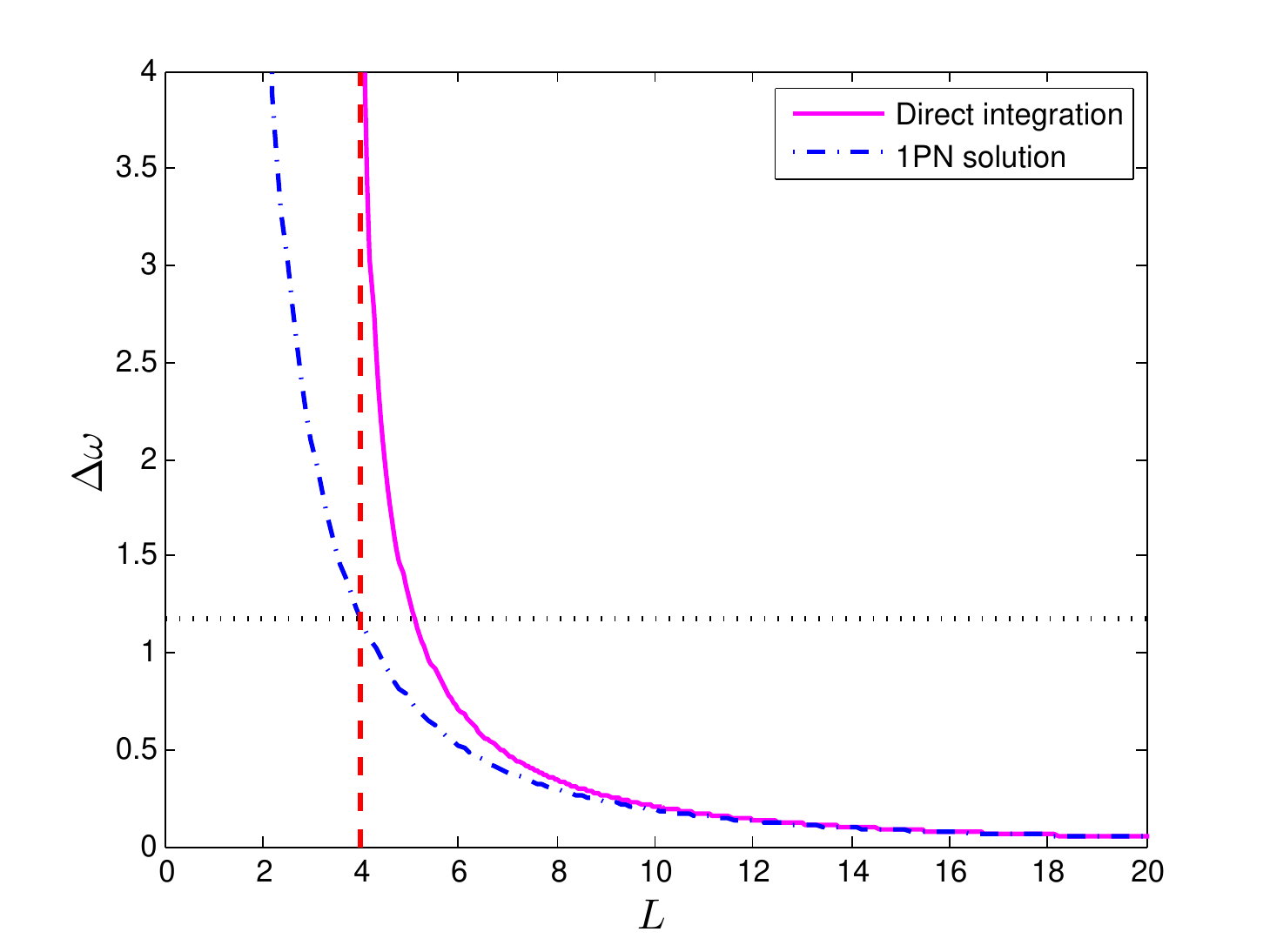}
\caption{Precession of argument of pericenter $\Delta \omega$ (in units of radians) as a function of the specific angular
momentum $L$.  The solid magenta and dot-dashed blue curves show the exact precession and its 1PN approximation, respectively.
The vertical dashed red line show the capture threshold $L_{\rm cap} = 4M_\bullet$ and the horizontal dotted line shows
the 1PN prediction for precession at this threshold.} \label{F:Dom}
\end{figure}

Early work on TDEs assumed that tidal debris would promptly circularize after falling back to pericenter
\cite{1988Natur.333..523R}, but Newtonian hydrodynamical simulations demonstrated that energy dissipation at pericenter
was inefficient for $M_\star \ll M_\bullet$ \cite{2014ApJ...783...23G}.  Early analytic work suggested that relativistic pericenter
precession might promote orbit circularization, because this precession would lead to steam crossings at which the inelastic
collisions of tidal elements would transform orbital kinetic energy into heat that could be subsequently radiated away.  This
suggestion was later supported by relativistic TDE simulations in which such precession was indeed shown to generate
tidal stream crossings \cite{2013MNRAS.434..909H,2013ApJ...775L...9D,2015ApJ...809..166G,2015ApJ...804...85S}.

The angular coordinate $\omega$ specifying the location of pericenter with respect to a reference axis in the orbital plane is
known as the argument of pericenter.  In Newtonian gravity, the argument of pericenter is a constant of motion, but this is
not true for Schwarzschild geodesics of the metric (\ref{E:Schw}) in Boyer-Lindquist coordinates.  At lowest post-Newtonian
(PN) order, the argument of pericenter changes by an amount
\begin{equation} \label{E:Dom1PN}
\Delta\omega_{\rm 1PN} =  6\pi \left( \frac{M_\bullet}{L} \right)^2
= 6\pi\beta \left( \frac{M_\bullet}{M_\star} \right)^{2/3} E_\star~.
\end{equation}
This PN approximation breaks down for $L \gtrsim L_{\rm cap}$ for which pericenter precession must be integrated
numerically along Schwarzschild geodesics,
\begin{equation} \label{E:DomS}
\Delta\omega_S = 2\int_r^\infty \frac{\dot{\phi}}{|\dot{r}|} dr' - 2\pi\, ,
\end{equation}
where
\begin{subequations}
\begin{align}
\dot{r} &= \pm \left\{ \frac{2M_\bullet}{r} \left[ 1 + \left( \frac{L}{r} \right)^2 \right] - \left( \frac{L}{r} \right)^2 \right\} \\
\dot{\phi} &= \left( \frac{L}{r} \right)^2
\end{align}
\end{subequations}
are the first-order derivatives of the Boyer-Lindquist coordinates $r$ and $\phi$ with respect to proper time $\tau$.  The
lower limit of the integral in Eq.~(\ref{E:DomS}) is the pericenter $r$ which depends implicitly on $L$ and hence $\beta$
through Eq.~(\ref{E:rpL}).  We can set the upper limit of this integral to $\infty$ since we assume the tidally disrupted stars
are initially on nearly parabolic orbits and the tidal debris remains highly eccentric.  Eq.~(\ref{E:relLPD}) also indicates that
we can approximate the specific angular momentum of the tidal debris as equal to that of the initial star.  We plot
$\Delta\omega_{\rm 1PN}$ and $\Delta\omega_S$ in Fig.~\ref{F:Dom}, which shows that pericenter precession diverges at
$L = L_{\rm cap}$ guaranteeing a tidal stream crossing and perhaps subsequent orbit circularization of the tidal debris.

\begin{figure}[t]
\centering
\includegraphics[width=0.52\textwidth]{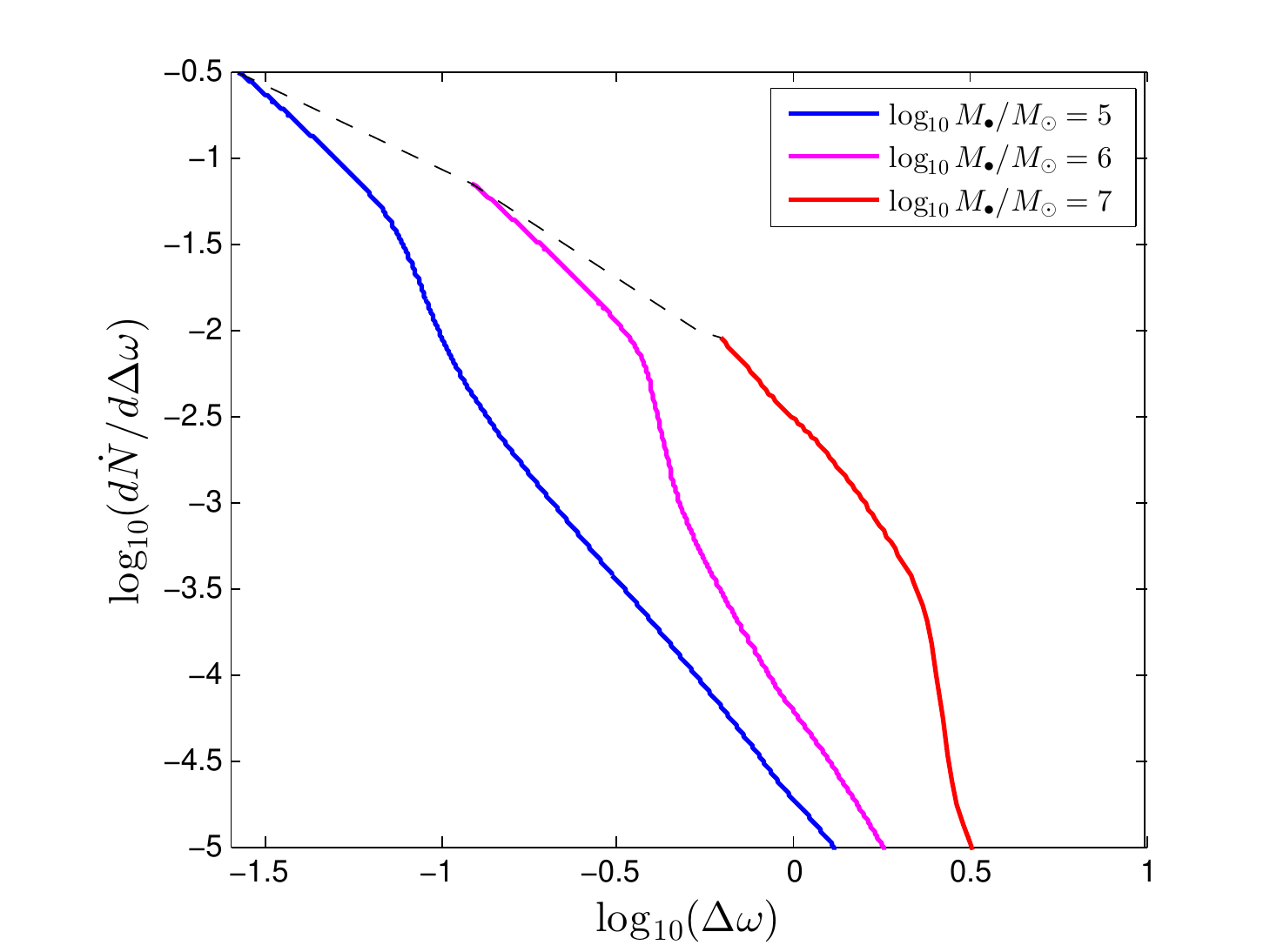}
\caption{The differential TDE rate $d\dot N/d\Delta\omega$ as a function of the precession of the argument of pericenter
$\Delta\omega$.  The solid blue, magenta, and red curves show this distribution for SBH masses
$M_\bullet/M_\odot = 10^5$, $10^6$, and $10^7$, respectively.  The dashed black curve connects the lower limit of each rate curve
corresponding to the threshold $\beta_{pd}$ of partial disruption.}
\label{F:dNdDom}
\end{figure}

Given that $\Delta\omega_S$ is a function of $\beta$, we can use the differential TDE rate $d\dot{N}/d\beta$ derived in
Sec.~\ref{SS:dbeta} and shown in the right panel of Fig.~\ref{F:dNdbeta} to derive the differential TDE rate
$d\dot{N}/d\Delta\omega$.  We show this rate in Fig.~\ref{F:dNdDom} for several SBH masses $M_\bullet$.  As expected
from Eq.~(\ref{E:Dom1PN}), TDEs by more massive SBHs have greater amount of pericenter precession suggesting that
relativistic orbit circularization might be more effective in such events.  Further investigation of this possibility requires a
criterion for orbit circularization that we hope to explore in future work.

\section{Discussion} \label{S:disc}

In this paper, we have systematically compared tidal disruption by Schwarzschild black holes in general relativity and
point masses in Newtonian gravity.  Differences between the two theories have potentially observable consequences for
both TDE rates and the properties of individual events:

(1) Tidal forces are stronger in general relativity than Newtonian gravity for orbits with the same penetration factor
$\beta \equiv L_t^2/L^2$.  This implies that the loss cone within which stars are tidally disrupted is larger in general relativity:
$L_d > L_{d,N} \Longrightarrow \beta_d < \beta_{d,N}$.  Partially disrupted stars with $\beta < \beta_d$ will lose more
material in relativity than Newtonian gravity, and stars with $\beta_d < \beta < \beta_{d,N}$ will be fully disrupted in
relativity but not Newtonian gravity.

(2) The width $\sigma_E$ of the energy distribution of tidal debris is larger in general relativity than Newtonian gravity for
orbits with the same penetration factor $\beta$.  This implies that stars will not only lose more material in partial disruptions
in relativity, but this material will become more tightly bound to the SBH and fall back to pericenter more quickly, leading to a
higher peak luminosity if the radiative efficiency is fixed.  However, stars that are fully disrupted in both theories
($\beta > \beta_{d,N}$) will be disrupted higher in the potential well in relativity than Newtonian gravity because of the
stronger tides.  This implies that the tidal debris will be less tightly bound despite the relativistic correction and therefore that
the peak fallback accretion rate is lower in relativity than Newtonian gravity.

(3) Black holes have event horizons in general relativity that allow them to directly capture stars.  This reduces the TDE rate
in relativity compared to Newtonian gravity, since tidal debris captured by the horizon cannot emit photons detectable by
observers.  As the threshold $L_{\rm cap}$ for direct capture increases more steeply with SBH mass than the thresholds
$L_d$ and $L_{pd}$ for full and partial disruptions, SBHs with masses greater than
$M_{\rm max, d} = 2.5 \times 10^7 M_\odot$ and $M_{\rm max, pd} = 1.4 \times 10^8 M_\odot$ will no longer be capable of
full and partial disruption, respectively.  These limits would have been underestimated by a factor of $\sim 4.5$ without
accounting for the stronger tides in relativity.

(4) Event horizons can capture a portion of the tidal debris in general relativity, even if the tidally disrupted star is not
initially on a capture orbit.  Tidal debris loses both energy and angular momentum during tidal disruption.  If the specific
angular momentum of the initial star was already close to the capture threshold, this additional loss can cause some of the
debris to plunge directly into the horizon.  This captured debris is the most tightly bound part of the tidal stream for
$\beta \lesssim \beta_d$, so its capture suppresses the early portions of the TDE light curve assuming the luminosity traces
the fallback accretion rate.

(5) Tidal streams precess in general relativity, potentially leading to inelastic collisions between parts of the stream that
allow energy to be dissipated and debris orbits to circularize.  At lowest PN order, this precession scales as
$M_\bullet^{2/3}$ at the threshold $\beta_d$ for full disruption, but it increases more steeply with SBH mass as $\beta_d$
approaches the threshold for direct capture $\beta_{\rm cap}$.

In future work, we plan to explore how SBH spin affects all five of these relativistic effects.  By breaking the spherical
symmetry of the spacetime, SBH spin increases the parameter space to include spin magnitude, orbital inclination, and
argument of pericenter.  As the tidal forces along geodesics depend on all these parameters, the diffusion equation
determining the rate at which stars enter the loss cone will likely need to be solved numerically to account for the
higher-dimensional boundary conditions.  Qualitatively, we expect TDE rates to be biased in favor of stars on retrograde
orbits for moderate SBH masses, but this bias should switch towards stars on prograde orbits as the retrograde
threshold for disruption falls below that for direct capture.  Both of these biases will tend to spin down the SBH, as a larger
fraction of the stellar material will be accreted in each case from retrograde orbits.  TDEs from stars on prograde orbits
however will be more luminous because of the more tightly bound prograde innermost stable circular orbits, leading to a
potential observational bias in favor of prograde TDEs.

SBH spin also affects relativistic precession.  At 1.5PN order, spin-dependent pericenter precession opposes (supports) the
1PN pericenter precession considered in this paper on prograde (retrograde) orbits \cite{2013degn.book.....M}, imposing a
further bias in favor of retrograde TDEs if such precession is required for orbit circularization.  SBH spin induces precession
of the longitude of ascending node which can cause tidal streams to precess out of their initial orbital planes, inhibiting
stream crossings and delaying orbit circularization \cite{2015ApJ...809..166G,2016MNRAS.461.3760H}.  Spin-dependent
solutions of stellar diffusion into the loss cone will allow us to calculate the fraction of TDEs experiencing such delays
between disruption and peak fallback accretion.  By providing a unified treatment of TDEs in the Schwarzschild spacetime,
this paper performs an important service in its own right and establishes a beachhead for attacking the more ambitious
problem of tidal disruption by spinning Kerr black holes.

\section*{Acknowledgements}

We would like to thank Chris Kochanek, David Merritt, and Nicholas Stone for useful conversations. M.~K. is supported by the Alfred P. Sloan Foundation Grant No. RG-2015-65299 and NSF Grant No. PHY-1607031.


\bibliography{Oct17}
\bibliographystyle{apsrev4-1}
\end{document}